# Data-Driven Modeling of an Unsaturated Bentonite Buffer Model Test Under High Temperatures Using an Enhanced Axisymmetric Reproducing Kernel Particle Method

Jonghyuk Baek[1], Yanran Wang[1], Xiaolong He[2], Yu Lu[1], John S. McCartney[1],

and J. S. Chen[1*]

[1] Department of Structural Engineering, University of California, San Diego, La Jolla, CA 92093, USA

[2] Ansys, Inc., Livermore, CA 94551, USA

## Abstract

In deep geological repositories for high level nuclear waste with close canister spacings, bentonite buffers can experience temperatures higher than 100 °C. In this range of extreme temperatures, phenomenological constitutive laws face limitations in capturing the thermo-hydro-mechanical (THM) behavior of the bentonite, since the pre-defined functional constitutive laws often lack generality and flexibility to capture a wide range of complex coupling phenomena as well as the effects of stress state and path dependency. In this work, a deep neural network (DNN)-based soil-water retention curve (SWRC) of bentonite is introduced and integrated into a Reproducing Kernel Particle Method (RKPM) for conducting THM simulations of the bentonite buffer. The DNN-SWRC model incorporates temperature as an additional input variable, allowing it to learn the relationship between suction and degree of saturation under the general non-isothermal condition, which is difficult to represent using a phenomenological SWRC. For effective modeling of the tank-scale test, new axisymmetric Reproducing Kernel basis functions enriched with singular Dirichlet enforcement representing heater placement and an effective convective heat

\* Corresponding author.

E-mail address: jsc137@ucsd.edu (J.S. Chen).

Postal address: 441H SME Building, University of California, San Diego, 9500 Gilman Drive, Mail Code 0085, La Jolla, CA 92093-0085

transfer coefficient representing thin-layer composite tank construction are developed. The proposed method is demonstrated through the modeling of a tank-scale experiment involving a cylindrical layer of MX-80 bentonite exposed to central heating.

# 1 Introduction

The safe and long-term disposal of radioactive waste is a major international challenge. The most common disposal approach under consideration is storing waste canisters in deep geological repositories (Pusch 1979). A key component in a geological repository is the engineered barrier system, consisting of a compacted bentonite buffer placed between the waste canister and the surrounding host rock (Lloret et al. 2003). Findings obtained from extended field studies on bentonite buffers suggest that the temperature, degree of saturation, and dry density distributions in the compacted bentonite buffer are subject to changes over the repository's operational duration (Villar et al. 2020). These changes arise due to the interaction between the heat released by the waste canister and the hydration driven by groundwater from the host rock. Consequently, accurately assessing and simulating the coupled thermal-hydro-mechanical behaviors of the compacted bentonite becomes essential for accurate estimation of the long-term buffer performance.

However, certain hurdles persist in developing proficient and accurate multi-physics models for bentonite buffers. One key challenge lies in the limited applicability of existing phenomenological constitutive models to accurately characterize the thermohydraulic properties of unsaturated bentonite under extremely high temperatures. These properties, including the soil-water retention curve (SWRC), the hydraulic conductivity function (HCF), the thermal conductivity function (TCF), and the volumetric heat capacity function (VHCF), are intricately coupled. This work focuses on the SWRC, given its influence on the other functional relationships. Specifically, Lu and McCartney (Lu and Mccartney 2023a) found that the HCF of compacted bentonite can be predicted from the shape of the SWRC and the parameters of the TCF correlate well with those of the SWRC (Lu and Dong 2015; Lu and McCartney 2023b). The VHCF has not been thoroughly studied but it is typically assumed to have similar parameters to the TCF (McCartney and Baser 2017). The SWRC is a vital relationship that describes unsaturated soil behaviors by connecting water content to matric suction (pore air and water pressure difference) (Fredlund and Rahardjo 1993). The SWRC is commonly estimated using empirical approaches. Various isothermal parametric models have been proposed in the literature, including the Brooks and Corey (BC) model (Brooks and Corey 1964), van Genuchten (VG) model (Genuchten 1980), and Fredlund and Xing (FX) model (Fredlund and Xing 1994). Several factors, such as the pore size distribution, porosity, pore fluid chemical composition, temperature, and soil mineralogy affect the SWRC (Villar and Lloret 2004), resulting in parametric models with multiple fitting parameters. Lu proposed (Lu 2016) a generalized SWRC model combining previous models that explicitly accounts for capillary and adsorptive water retention mechanisms, which is shown to be a statistical improvement over other models, especially for high plasticity clays like bentonite. Efforts have been made to integrate temperature effects into SWRC parametric models (Zhou et al. 2014; Wan et al. 2015; Roshani and Sedano 2016), but these only considered the interfacial surface tension as the sole temperature-dependent variable, neglecting other parameters. Vahedifard et al. (Vahedifard et al. 2018) introduced non-isothermal extensions of BC, VG, and FX models accounting for the temperature effects on surface tension, the contact angle, and adsorption. The proposed non-isothermal extensions are validated against experimental data under various temperatures, but the number of fitting parameters increases compared to the original isothermal models.

In parametric-based estimation of the SWRC, defining accurate relationships between textural data and hydraulic properties a priori is crucial. Yet, this task is non-trivial due to the typically unknown nature of these relationships. Recent progress in machine learning and data science have brought great opportunities to develop data-driven constitutive modeling (Vlassis et al. 2020; Vlassis and Sun 2021; Xu et al. 2021; He and Chen 2022; He and Semnani 2023; Xiong et al. 2023), data-driven computational mechanics (Kirchdoerfer and Ortiz 2016; Ibañez et al. 2018; Eggersmann et al. 2019; He and Chen 2020; He et al. 2020; He et al. 2021; Bahmani and Sun 2023), multiscale modeling (Liu et al. 2016; Wang and Sun 2018; He et al. 2022; Bishara et al. 2023; Wei et al. 2023), reduced-order modeling (Qian et al. 2020; Kaneko et al. 2021; Kim et al. 2022; Fries et al. 2022; Issan and Kramer 2023; He et al. 2023; Bonneville et al. 2023), system or parameter identification (Peherstorfer and Willcox 2016; Brunton et al. 2016; Raissi et al. 2019; Cranmer et al. 2020; Kadeethum et al. 2021; Taneja et al. 2022; Taneja et al. 2023), damage and fracture modeling (Baek et al. 2022; Baek and Chen 2023), etc., due to their strong flexibility and representation capability (Hornik 1991; Goodfellow et al. 2016). Pachepsky et al. (Pachepsky et al. 1996) applied artificial neural networks (ANNs) to learn how soil texture and bulk density are related to hydraulic properties of soil using experimental data from two types of soil samples at eight matric potentials. The authors compared the performance of the proposed ANN model with polynomial regressions combined with the VG model and found that the ANN is on par with polynomial regressions with carefully fitted coefficients, indicating its ability to estimate soil hydraulic properties from easily measurable soil data, such as soil texture and density. Koekkoek and Booltink (Koekkoek and Booltink 1999) designed and evaluated three deep neural networks (DNNs) with varying input variables, including soil texture and bulk density as considered in (Pachepsky et al. 1996), and additional variables, such as matric potential and ped-size. All DNNs exhibited performance enhancements compared to regression models, especially the one with the most input variables, highlighting DNNs' efficacy in expressing water retention dependencies. Using genetic programming, Johari et al. (Johari et al. 2006) devised a model with input terminals consisting of soil characteristics, initial conditions, and normalized suctions, a functional set containing algorithmic operators, and an output terminal for gravimetric water content. A functional form describing the water retention behaviors was generated from an experimental database with 2694 patterns, showcasing better SWRC prediction performance compared to traditional parametric models. However, the proposed approach necessitates extensive datasets from a relatively large number of experiments. Zhang et al. (Zhang et al. 2022) introduced a hybrid approach fusing Bayes' theorem with prior knowledge from the FX model. The Bayesian model utilizes Markov Chain Monte Carlo simulation to update uncertain variables based on limited measured data. The proposed approach offered accurate SWRC estimation with sparse experimental data and permitted quantification of the impact of measured points on SWRC estimation with careful selection of measured points that is crucial for effective uncertainty reduction. Notably, temperature dependency of the SWRC is not addressed in the machine learning methods mentioned above.

In this work, a non-isothermal, data-driven, DNN-based SWRC model is introduced. Since this work only targets the MX-80 bentonite, the soil characteristics and compositions are not considered as inputs for the machine learning model. The DNN model takes the temperature and matric suction as inputs and predicts degrees of saturation. Sparse experimental data sourced from (Villar and Gómez-Espina 2008) and (Villar and Gòmez-Espina 2007) concerning bentonite's suction evolutions during heating at five levels of gravimetric water contents are utilized for model training. To expand the training dataset, the experimental data is used to fit separate curves of Lu's isothermal SWRC model at different temperatures ranging from 24 to 100 ℃. The trained DNN-SWRC model then serves as a constitutive model for bentonite's water retention behaviors.

Another challenge emerges in developing reliable and effective numerical techniques to model the highly coupled thermo-hydro-mechanical (THM) behavior within the bentonite buffers. Many computer codes have been developed for modeling coupled THM processes, such as ROCMAS (Noorishad and Tsang 1996), FRACON (Nguyen 1996), ABAQUS (Börgesson 1996), COMPASS (Thomas et al. 1996), CODE-BRIGHT (Olivella et al. 1994), and TOUGH-FLAC (Zheng et al. 2015). These codes employ simplifications like assuming constant low gas pressure, steady porosity, and a basic bulk strain function. While primary unknown variables are reduced from these simplifications, they might disrupt phase consistency and numerical stability. Tong et al. (Tong et al. 2010) presented a numerical approach to simulate coupled THM processes in porous geomaterials, considering solid phase deformation, multiphase fluid flow, and heat transport based on the mixture theory. Their model was validated with an experimental test of a cylindrical, compacted MX-80 bentonite sample experiencing heating and gradual hydration. Although the model was proved to be comprehensive and reliable, challenges remain due to reliance on empirical thermal-hydraulic property relations, which might lack physical reliability. Wei et al. (Wei et al. 2016) proposed a fully coupled, stabilized meshfree formulation for hydro-mechanical analysis in fluid-saturated porous media. It employed an equal-ordered $\mathbf{u} - p$ reproducing kernel approximation with fluid pressure projection stabilization, which effectively addresses non-physical spatial oscillations in fluid pressure under nearly impermeable or undrained conditions. Xie and Wang (Xie and Wang 2014) employed an iterative coupling scheme with reproducing kernel particle method (RKPM) to enhance stability and avoid potentially ill-conditioned global matrices resulted from fully coupled systems in solving coupled hydro-mechanical problems. They utilized an equal-ordered $\mathbf{u} - p$ formulation with a stabilization technique, proving unconditional stability with an optimal relaxation parameter.

In this study, we introduce a meshfree multi-physics formulation based on RKPM for coupled thermo-hydraulic modeling under high-temperature conditions to simulate complex interactions between heat transfer and water flows in the bentonite buffers. An enriched function tailored for axisymmetric diffusion problems is developed and incorporated into the RKPM, allowing for accurate representation of sharp temperature transitions near the heat source. A semi-analytical treatment is proposed for the convection boundary with thin composite layers, where an effective convection transfer coefficient is introduced to mitigate excessive spatial and temporal discretization in the tank's thin outer layers. To enhance numerical stability and efficiency, the stabilized conforming nodal integration (SCNI) (Chen et al. 2001) is reformulated for the axisymmetric geometry. The proposed meshfree numerical approach is integrated with the DNN-SWRC constitutive model for a tank-scale simulation of heating and gradual dehydration of a cylindrical, compacted MX-80 bentonite specimen.

The remainder of the paper is organized as follows. The experimental setups of the tank-scale test on compacted MX-80 bentonite are described in Section 2. Section 3 provides basic equations for the mechanics of the coupled thermo-hydraulic problem and an overview of the reproducing kernel particle method. In Section 4, enhanced RKPM formulations are introduced, including axisymmetric enrichment, thin-layer convection boundary treatment, and the reformulated SCNI for the axisymmetric geometry, along with validation problems. Section 5 presents the computational graph and the training procedure of a non-isothermal, data-driven, DNN-based SWRC constitutive model. Comparisons of performances between the trained DNN-SWRC model and Lu's SWRC model are also provided. In section 6, numerical setups of a tank-scale experiment, as well as numerical simulation results are illustrated, and the paper concludes with a discussion and summary in Section 7.

## 2  Tank-Scale Experiment on Compacted Bentonite

This work focuses on simulating the coupled heat transfer and hydration in a layer of compacted MX-80 bentonite during central heating reported in the study by Lu and McCartney (Lu and McCartney 2022). A cross-sectional schematic of the tank scale experimental setup is shown in Figure 2-1. The bentonite layer was compacted within a cylindrical aluminum testing tank having an inner radius of 277.3 mm, and a layer of fiberglass was wrapped around the testing tank for insulation. A cartridge heater with a diameter of 12.5 mm is installed at the center of the bentonite layer, and five dielectric sensors are placed at 50, 70, 100, 125, and 185 mm radial distances away from the surface of the heater to provide experimental measurements of the solid temperature and water content. Relative humidity sensors were also included in the setup but are not evaluated in this study. The room temperature, the temperatures at the top of the soil layer above the heater and at the inside edge of the tank are also monitored by the thermocouples. A reinforced concrete cap was placed above the bentonite layer to provide confinement, but the cap was not restrained from moving in the heating stage of this experiment. A linearly variable differential transformer (LVDT) is used to measure the vertical displacement of the cap. The initial condition of the compacted MX-80 bentonite soil layer has a height of 210.5 mm, and the bentonite was compacted to an initial porosity of 0.42 and an initial degree of saturation of 0.32. Further details of the bentonite material properties and testing procedures can be found in Lu and McCartney (Lu and McCartney 2022; Lu and Mccartney 2023a).

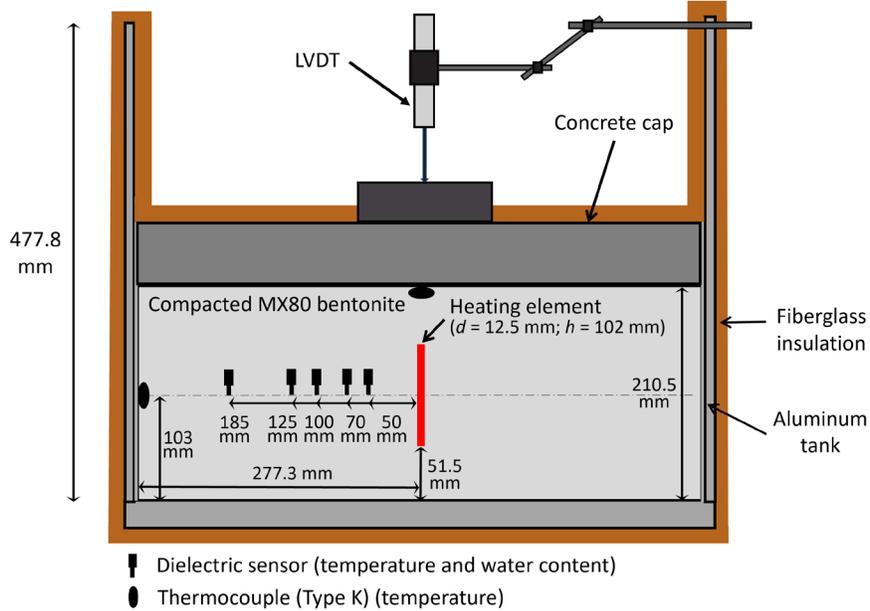

**Figure 2-1: A cross-sectional schematic of the tank scale experimental setup**
(recreated from Figure 3 (b) in (Lu and McCartney 2022))

## 3 Basic Equations

### 3.1 Mechanics of the problem

Let $\Omega$ be a domain of a triphasic porous body that is represented as a continuum of an immiscible mixture of solid and water, the physical quantities of which are determined by spatial averaging. Let $\boldsymbol{u}(\mathbf{x}, t)$, $T(\mathbf{x}, t)$, and $p(\mathbf{x}, t)$ denote the solid displacement, temperature, and water pressure of a position $\mathbf{x} \in \Omega$ at a time $t$. Because the primary goal of this work is to numerically simulate the thermo-hydraulic interactions of bentonite in the heating stage of the tank scale experiments without water inflow, only soil temperature and water pressure are considered as independent variables as negligible axial and radial displacements are observed in the experiments and the air pressure is assumed to be constant and equal to atmospheric pressure to replicate the conditions simulated in the tank-scale test.

To model the thermo-hydraulic behavior of bentonite, two coupled balance equations need to be considered. The first one is the fluid mass conservation equation, which is formulated as follows:

$$\frac{d(\phi S \rho^w)}{dt} + \nabla \cdot (\rho^w \mathbf{q}^w) = 0, \tag{1}$$

where $\phi$ denotes the soil porosity, and $\mathbf{q}^w$ is the water seepage velocity. The water seepage velocity is defined as follows for the Darcian flow:

$$\mathbf{q}^w = -k^w(\nabla p - S\rho^w \mathbf{g}) - k^o \nabla T, \tag{2}$$

where $k^w$, $S$, $\rho^w$, $k^o$, and $\mathbf{g}$ represent the hydraulic conductivity, degree of saturation, water density, thermo-osmotic permeability, and the gravitational acceleration, respectively. The first term of the left-hand side of Eq. (1) can be expanded using the product rule:

$$S\rho^w \frac{d\phi}{dt} + \phi \rho^w \frac{dS}{dt} + \phi S \frac{d\rho^w}{dt} + \nabla \cdot (\rho^w \mathbf{q}^w) = 0. \tag{3}$$

Porosity depends on the compaction and deformation of the solid. Take the form provided in (Sun 2015) and assuming negligible deformation, the time rate change of porosity can be written as:

$$\frac{d\phi}{dt} = -(1 - \phi_0)\alpha_s \frac{dT}{dt}, \tag{4}$$

where $\alpha_s$ is the thermal expansion coefficient and $\phi_0$ represents the initial porosity. The degree of saturation is both temperature and fluid pressure dependent, so its time derivative can be rewritten from the chain rule:

$$\frac{dS}{dt} = \frac{\partial S}{\partial p}\frac{dp}{dt} + \frac{\partial S}{\partial T}\frac{dT}{dt} = -\frac{\partial S}{\partial \psi}\frac{dp}{dt} + \frac{\partial S}{\partial T}\frac{dT}{dt}, \tag{5}$$

where $\psi$ denotes the matric suction and is related to the water pressure as $\psi = -p$, assuming negligible air pressure. The water density is assumed to be temperature dependent (Hillel 1980) so its time derivative can be expanded as follows:

$$\frac{d\rho^w}{dt} = \frac{\partial \rho^w}{\partial T}\frac{dT}{dt}. \tag{6}$$

Substitute the time derivatives shown in Eqs. (4)-(6) into Eq. (3), and assume $\phi \approx \phi_0$ for restrained conditions (constant volume), the fluid mass conservation equation can be rearranged as:

$$\phi_0 \rho^w \frac{\partial S}{\partial p} \dot{p} + \left( \phi_0 \rho^w \frac{\partial S}{\partial T} + \phi_0 S \frac{\partial \rho^w}{\partial T} - S \rho^w (1 - \phi_0) \alpha_s \right) \dot{T} + \nabla \cdot (\rho^w \mathbf{q}^w) = 0, \quad (7)$$

where $(\dot{\cdot})$ denotes the first order time derivative.

The second equation is the transient heat equation, which is expressed as follows:

$$C_v \dot{T} + \nabla \cdot \mathbf{q} + \phi_0 S C_v^w \mathbf{q}^w \cdot \nabla T = 0, \quad (8)$$

where $C_v$ and $C_v^w$ denote the volumetric heat capacity of the solid and water, respectively, and $\mathbf{q}$ is the heat flux density, which is represented by Fourier's law:

$$\mathbf{q} = -k^T \nabla T, \quad (9)$$

where $k^T$ is the thermal conductivity.

Along with the above equations, the strong form of the problem involves the following boundary conditions:

$$\begin{aligned} T &= \bar{T} & \text{on } \Gamma_g, \\ \bar{\mathbf{q}}^{conv} &= \bar{c}(T - T^{env}) & \text{on } \Gamma_t, \\ p &= \bar{p} & \text{on } \Gamma_p, \\ \mathbf{n}_q \cdot \mathbf{q}^w &= \mathbf{n}_q \cdot \bar{\mathbf{q}}^w & \text{on } \Gamma_q, \end{aligned} \quad (10)$$

and the following initial conditions:

$$T = T_0, \quad p = p_0, \quad \text{at } t = 0, \quad (11)$$

where $\bar{T}$ is the applied temperature on the solid temperature Dirichlet boundary $\Gamma_g$, $\Gamma_t$ denotes the effective convective boundary with an effective convective coefficient $\bar{c}$, $\bar{p}$ is the applied fluid pressure on the fluid pressure boundary $\Gamma_p$, $\bar{\mathbf{q}}^w$ and $\mathbf{n}_q$ are the applied water seepage velocity and the outward normal on the fluid inflow boundary $\Gamma_q$, respectively, and $\Gamma_g \cup \Gamma_t = \Gamma$, $\Gamma_p \cup \Gamma_q = \Gamma$, $\Gamma_g \cap \Gamma_t = \emptyset$, $\Gamma_p \cap \Gamma_q = \emptyset$. As shown from two balance equations in Eq. (7) and Eq. (8), the unknown fluid pressure and solid temperature are fully coupled. In addition, these balance equations involve coupled thermo-hydraulic properties of bentonite: the SWRC which relates the degree of saturation with suction, the HCF that relates the degree of saturation and hydraulic conductivity, the TCF that relates degree of saturation and the thermal conductivity and the VHCF that relates the degree of saturation and the volumetric heat capacity. The SRWC and HCF may be temperature-dependent (Lu and Mccartney 2023a) while it is typically assumed that the TCF and VHCF are not temperature dependent.

### 3.1.1 Galerkin formulation

To formulate the weak form of two coupled balance equations, two spaces of trial functions for both solid temperature and fluid pressure fields are defined as:

$$S_T = \{T: \Omega \to \mathbb{R} | T \in H^1, T = \bar{T} \text{ on } \Gamma_g\}, \quad (12)$$
$$S_p = \{p: \Omega \to \mathbb{R} | p \in H^1, p = \bar{p} \text{ on } \Gamma_p\}.$$

Then the corresponding spaces for the test functions for $T$ and $p$ are:

$$V_T = \{\delta T: \Omega \to \mathbb{R} | \delta T \in H^1, \delta T = 0 \text{ on } \Gamma_g\}, \quad (13)$$
$$V_p = \{\delta p: \Omega \to \mathbb{R} | \delta p \in H^1, \delta p = 0 \text{ on } \Gamma_p\}.$$

The weak form statement of the coupled balance equations is then to find $(T, p) \in S_T \times S_p$ such that for all $(\delta T, \delta p) \in V_T \times V_p$,

$$\int_\Omega \delta p \phi_0 \rho^w \frac{\partial S}{\partial p} \dot{p} d\Omega$$
$$+ \int_\Omega \delta p \left( \phi_0 \rho^w \frac{\partial S}{\partial T} + \phi_0 S \frac{\partial \rho^w}{\partial T} - S \rho^w (1 - \phi_0) \alpha_s \right) \dot{T} d\Omega \quad (14)$$
$$- \int_\Omega \boldsymbol{\nabla} \delta p \cdot (\rho^w \boldsymbol{q}^w) d\Omega + \int_{\Gamma_q} \delta p \rho^w \mathbf{n}_q \cdot \bar{\mathbf{q}}^w d\Gamma = 0,$$

$$\int_\Omega \delta T C_v \dot{T} d\Omega + \int_\Omega \boldsymbol{\nabla} \delta T k^T \boldsymbol{\nabla} T d\Omega + \int_{\Gamma_t} \delta T \bar{c}(T - T^{env}) d\Gamma \quad (15)$$
$$+ \int_\Omega \delta T \phi_0 S C_v^w \boldsymbol{q}^w \cdot \boldsymbol{\nabla} T \, d\Omega = 0.$$

The linearization of weak formulations of two coupled equations can be obtained by the standard linearization procedures (Belytschko et al. 2013).

Let $T^h$ and $p^h$ be approximations of the solid temperature $T$ and the water pressure $p$, respectively, and let both variables be approximated by the reproducing kernel approximation as follows:

$$T^h(\mathbf{x},t) = \sum_{I=1}^{NP} \bar{\Psi}_I(\mathbf{x})\tau_I(t) \equiv \bar{\boldsymbol{\Psi}}\boldsymbol{\tau}, \qquad (16)$$

$$p^h(\mathbf{x},t) = \sum_{I=1}^{NP} \Psi_I(\mathbf{x})d_I(t) \equiv \boldsymbol{\Psi}\mathbf{d}, \qquad (17)$$

where $\Psi_I$ is the regular RK shape function of node $I$, and $\bar{\Psi}_I$ is the enriched RK shape function, which will be introduced in Section 4.1, and $\tau_I$ and $d_I$ denote the associated approximation coefficients of node $I$ to be sought corresponding to the temperature and pressure fields, respectively. Furthermore, let $\delta T^h$ and $\delta p^h$ take the same form of approximations as their trial function counterparts:

$$\delta T^h(\mathbf{x}) = \sum_{I=1}^{NP} \bar{\Psi}_I(\mathbf{x})\delta\tau_I, \qquad (18)$$

$$\delta p^h(\mathbf{x}) = \sum_{I=1}^{NP} \Psi_I(\mathbf{x})\delta d_I. \qquad (19)$$

The statement of the Galerkin formulation of the water mass balance equation is to seek the approximated solution fields $(T^h, p^h) \in S_T^h \times S_p^h \subset S_T \times S_p$, so that for all variations of $(\delta T^h, \delta p^h) \in V_T^h \times V_p^h \subset V_T \times V_p$,

$$\begin{aligned}
&\int_\Omega \delta p^h \phi_0 \rho^w \left(\frac{\partial S}{\partial p}\right)^h \dot{p}^h d\Omega \\
&+ \int_\Omega \delta p^h \left(\phi_0 \rho^w \left(\frac{\partial S}{\partial T}\right)^h + \phi_0 S^h \left(\frac{\partial \rho^w}{\partial T}\right)^h - S^h \rho^w (1-\phi_0)\alpha_s\right) \dot{T}^h d\Omega \\
&- \int_\Omega \boldsymbol{\nabla}\delta p^h \cdot (\rho^w \mathbf{q}^{wh}) d\Omega + \int_{\Gamma_q} \delta p^h \rho^w \mathbf{n}_q \cdot \bar{\mathbf{q}}^w d\Gamma = 0,
\end{aligned} \qquad (20)$$

$$\begin{aligned}
&\int_\Omega \delta T^h C_v \dot{T}^h d\Omega + \int_\Omega \boldsymbol{\nabla}\delta T^h k^T \boldsymbol{\nabla} T^h d\Omega + \int_{\Gamma_t} \delta T^h \bar{c}(T^h - T^{env}) d\Gamma \\
&+ \int_\Omega \delta T^h \phi_0 S^h C_v^w \mathbf{q}^{wh} \cdot \boldsymbol{\nabla} T^h d\Omega = 0.
\end{aligned} \qquad (21)$$

## 3.2 Overview of Reproducing Kernel Particle Method (RKPM)

In the reproducing kernel (RK) approximation (Liu et al. 1995; Chen et al. 1996; Chen et al. 2017), the RK shape functions are constructed such that a certain set of monomial basis functions and physics-driven characteristic functions are exactly reproduced, providing flexibility in constructing a tailored approximation space suitable for a particular problem under consideration. This property allows the simulation models to efficiently capture critical solution behavior such as a sharp temperature transition near the heat element in the tank scale test simulated in this study.

### 3.2.1 Reproducing kernel approximation

Let a closed domain $\bar{\Omega} = \Omega \cup \partial\Omega \subset \mathbb{R}^d$ be discretized by a set of $NP$ nodes with nodal coordinate $\mathbf{x}_I$, $1 \leq I \leq NP$, as shown in Figure 3-1. The RK approximation $f^h(\mathbf{x})$ of a function $f(\mathbf{x})$ is:

$$f^h(\mathbf{x}) = \sum_{I=1}^{NP} \Psi_I(\mathbf{x}) d_I, \tag{22}$$

where $\Psi_I(\mathbf{x})$ is RK shape function with a compact support centered at node $I$ and $d_I$ is the generalized nodal coefficient of node $I$ as described in Figure 3-1. $\Psi_I(\mathbf{x})$ is constructed such that the following set of discrete reproducing conditions is satisfied (Chen et al. 1998):

$$\sum_{I=1}^{NP} \Psi_I(\mathbf{x})(\mathbf{x} - \mathbf{x}_I)^{\boldsymbol{\alpha}} = \delta_{0\boldsymbol{\alpha}}, \quad |\boldsymbol{\alpha}| \leq n,$$

or

$$\sum_{I=1}^{NP} \Psi_I(\mathbf{x}) \mathbf{H}(\mathbf{x} - \mathbf{x}_I) = \mathbf{H}(\mathbf{0}), \tag{23}$$

where $\boldsymbol{\alpha} = (\alpha_1, \alpha_2, \ldots, \alpha_d)$ is a multi-index notation with a length defined as $|\boldsymbol{\alpha}| \equiv \sum_{i=1}^{d} \alpha_i$, and $\mathbf{x}^{\boldsymbol{\alpha}} \equiv x_1^{\alpha_1} \cdot x_2^{\alpha_2} \cdot \ldots \cdot x_d^{\alpha_d}$. By setting $\Psi_I(\mathbf{x}) = \mathbf{H}^T(\mathbf{x} - \mathbf{x}_I)\mathbf{b}(\mathbf{x})$ and solving Eq. (23) for $\mathbf{b}(\mathbf{x})$, the following RK shape function is obtained:

$$\Psi_I(\mathbf{x}) = \mathbf{H}^T(\mathbf{0}) \mathbf{M}^{-1}(\mathbf{x}) \mathbf{H}(\mathbf{x} - \mathbf{x}_I) \Phi_a(\mathbf{x} - \mathbf{x}_I), \tag{24}$$

where the moment matrix $\mathbf{M}(\mathbf{x})$ and the basis vector $\mathbf{H}(\mathbf{x} - \mathbf{x}_I)$ are defined as:

$$\mathbf{M}(\mathbf{x}) = \sum_{I=1}^{NP} \mathbf{H}(\mathbf{x} - \mathbf{x}_I)\mathbf{H}^T(\mathbf{x} - \mathbf{x}_I)\Phi_a(\mathbf{x} - \mathbf{x}_I), \quad (25)$$

$$\mathbf{H}(\mathbf{x} - \mathbf{x}_I) = [1, (x_1 - x_{1I}), (x_2 - x_{2I}), \cdots, (x_3 - x_{3I})^n]^T, \quad (26)$$

and $\Phi_a(\mathbf{x} - \mathbf{x}_I)$ is the kernel function with a compact support "$a$", such as a cubic B-spline function. The kernel function controls the locality and smoothness (or roughness) or approximated solutions, while the orders of approximation are determined by the order of basis function $n$. Note that an additional set of reproducing conditions can be introduced to capture certain features of the problems of interest without adding additional degrees of freedom to the approximation. This flexibility is utilized in this work to capture a sharp temperature transition behavior as described in Section 4.1.

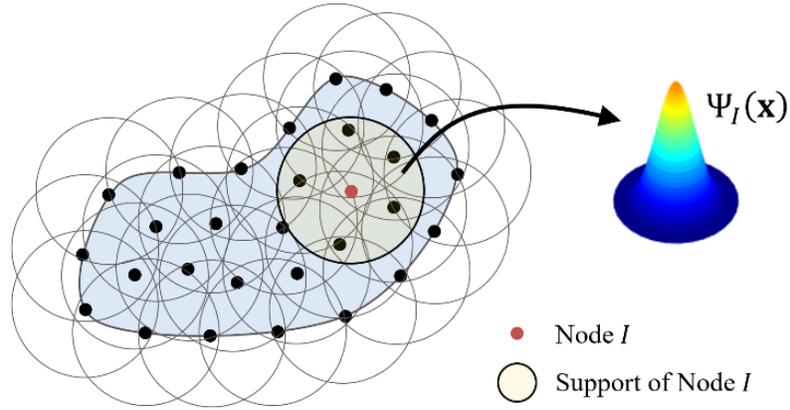

Figure 3-1: Illustration of RK discretization and shape function

### 3.2.2 Stabilized Conforming Nodal Integration (SCNI)

When Gauss integration on background integration cells (Dolbow and Belytschko 1999) is used for numerical integration of RK approximation in the Galerkin weak form, a significantly high-order rule is required to yield optimal solution convergence (Chen et al. 2017), due to the rational shape function given in Eq. (24). This, in turn, leads to a significant increase in computational cost. To address this issue, the stabilized conforming nodal integration (SCNI) was proposed in (Chen et al. 2001).

In SCNI, the domain is partitioned into $N_{IC}$ conforming smoothing cells for numerical integration, such as Voronoi cells, as illustrated in Figure 3-2 where $N_{IC}$ denotes the number of smoothing cells. The SCNI method is formulated to exactly meet linear variational consistency between the variational formulation and the partial differential equation and also remedy the rank deficiency in the direct nodal integration method (Beissel and Belytschko 1996) by introducing the following smoothed gradient in the Galerkin approximation:

$$\widetilde{\nabla}\Psi_I(x_L) = \frac{1}{W_L}\int_{\Omega_L}\nabla\Psi_I(x)\mathrm{d}\Omega = \frac{1}{W_L}\int_{\partial\Omega_L}\Psi_I(x)\boldsymbol{n}\,\mathrm{d}\Gamma, \qquad W_L = \int_{\Omega_L} d\Omega, \qquad (27)$$

where $\Omega_L$ denotes the nodal representative conforming smoothing cells, and $\boldsymbol{n}$ represents the unit outward normal of the smoothing cell boundary $\Gamma_L$. It has been proved that SCNI passes linear patch tests with single point integration on each boundary and yields optimal convergence rates with high accuracy for multi-dimensional problems (Chen et al. 2001; Chen et al. 2013).

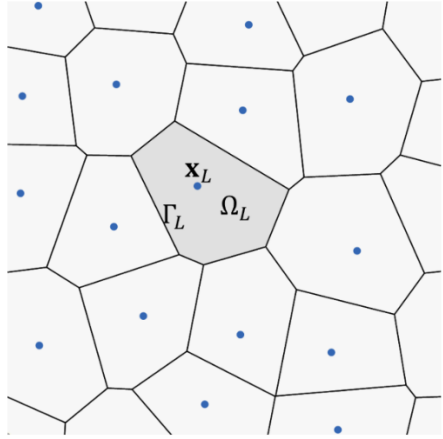

Figure 3-2: Conforming integration cell used in SCNI: $\Omega_L$, $\Gamma_L$, and $x_L$ denote the domain, the boundary, and the centroid of the integration cell L, respectively.

Compared to high-order Gauss integration, SCNI is computationally much more efficient as it eliminates the need to evaluate direct derivatives of RK shape functions at a large number of integration points. Note that while $N_{IC}$ coincides with the number of particles for standard meshfree methods, the smoothing cells can be adaptively refined to improve accuracy.

# 4 Enhanced Axisymmetric RK Approximation

## 4.1 RK approximation for axisymmetric heat transfer problems

An axisymmetric heat transfer problem with prescribed temperature (such as the heater in a repository or in a tank-scale test) at the line of axisymmetry exhibits a high gradient characteristic near the surface of prescribed temperature. In the limit of prescribed temperature location approaching the line of symmetry, the temperature gradient approaches infinity, posing difficulties in numerical simulation. For instance, when a heater is placed at the line of axisymmetry in the tank scale experiments of bentonite, the soil exhibits sharp temperature drops within 50 mm from the center of the tank. This requires high-resolution discretization in traditional methods and results in high computational costs. To effectively capture the sharp temperature gradient near the heater element, an enriched basis function for RKPM temperature approximation has been formulated based on the derived analytical solution of a radial plane-strain heat conduction boundary value problem in the problem domain $\Omega = [R_i, R_o]$, with prescribed temperature $T = T_0$ at $r = R_i$, and convection boundary at $r = R_o$ with heat transfer coefficient $c$ and environmental temperature $T^{env}$:

$$T = -\frac{A}{k}\log(r) + B, \tag{28}$$

where $T$, $k^T$, and $r$ are temperature, thermal conductivity, and radial coordinate, respectively. The constant $A$ and $B$ are:

$$A = \frac{cR_o(T_0 - T^{env})}{1 + (cR_o/k)\log(R_o/R_i)},$$

$$B = T_0 + \frac{A}{k}\log(R_i). \tag{29}$$

It is easily shown that $dT/dr \to \infty$ as $R_i \to 0$. The enriched RK basis in the cylindrical coordinate is constructed by considering the following additional reproducing condition:

$$\sum_I \Psi_I(r) \log(r_I) = \log(r) \qquad (30)$$

$$= \left(\sum_I \Psi_I(r)\right) \log(r) \rightarrow \sum_I \Psi_I(r) \log(r/r_I) = 0,$$

which leads to the following enriched basis vector for the RK approximation:

$$\bar{\mathbf{H}} = [1 \quad r - r_I \quad z - z_I \quad \log(r/r_I)]^T. \qquad (31)$$

The partition of unity of the RK shape functions, i.e., $\sum_I \Psi_I(r) = 1$, is utilized in Eq. (30). Note that the enriched RK approximation with combined monomial and logarithm basis functions in Eq. (31) is applicable to solution at any location in the axisymmetric domain with or without gradient singularities. The coefficients of these basis functions are obtained as part of the Galerkin-based solution of the heat transfer equation without introducing additional degrees of freedom. For problems without gradient singularities or for locations far away from the line of axisymmetry, the logarithm term can be dropped for better computational efficiency.

A radial-plane strain heat transfer problem is considered to demonstrate the effectiveness of the proposed RK basis enrichment. Figure 4-1 shows the problem settings, where a prescribed temperature of 200 °C is set at the inner boundary and an effective convection boundary condition is applied on the outer boundary with thermal conductivity of 0.46 W/m/K, effective convective heat transfer coefficient of 2.008 $\text{Wm}^{-2}\text{K}^{-1}$, and a room temperature of 22 °C. As shown in Figure 4-2 (a), the un-enriched RKPM only offer marginal accuracy compared to the analytical solution even with very fine nodal spacing ($h = 1/800$ of the tank radial dimension), while all the enriched RKPM cases achieve good accuracy, as illustrated in Figure 4-2 (b), even with very coarse discretization ($h = 1/100$ of the tank radial dimension).

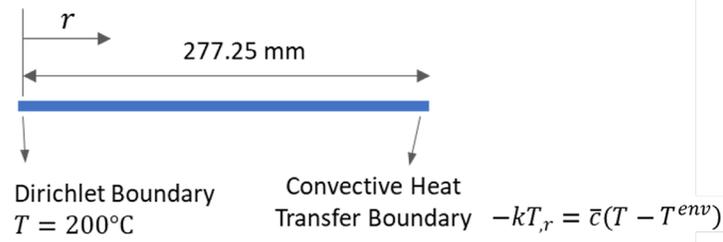

Figure 4-1: Heat transfer problem settings

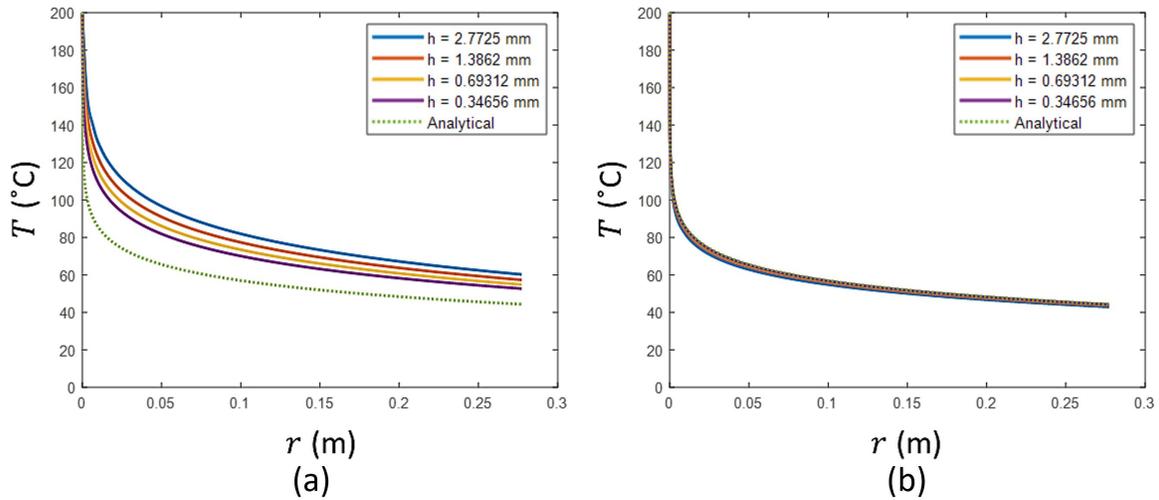

Figure 4-2: (a) un-enriched RKPM and analytical solutions and (b) the enriched RKPM and analytical solutions. (In the legends, $h$ denotes the nodal spacing used at the region around $r = 0$, which is the axis of axisymmetry)

## 4.2 Semi-analytical treatments for the convection boundary with thin composite layers

To introduce a realistic numerical model for the tank scale experiments, the outer aluminum tank thin layer and the fiberglass insulation layer, as shown in Figure 2-1, need to be considered, in addition to the soil material. However, the aluminum tank and fiberglass layers are much thinner compared to the soil specimen, demanding dense spatial discretization in the numerical model, and correspondingly the small stable time step size for time integration. To resolve this issue, a semi-analytical approach is taken to obtain an "effective heat transfer coefficient" $\bar{c}$, representing the effective convective heat transfer through the outer double-layer materials (aluminum and fiberglass layers) without explicitly modeling those thin layers. Figure 4-3 schematically describes a radial plane-strain full model with the outer layers and its corresponding reduced-order model without the outer layers while subjected to an "effective" convective heat transfer.

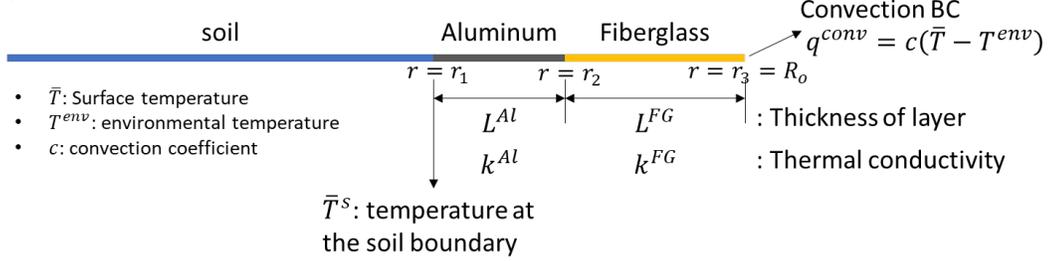

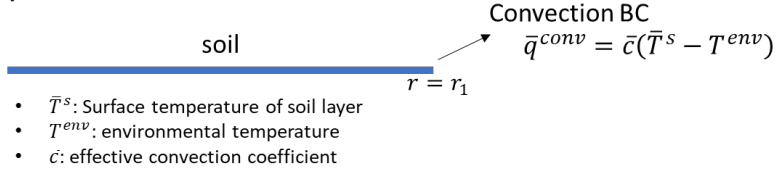

**Figure 4-3: Schematic description of radial plane-strain models: (a) full model with the outer layers and (b) reduced model (without the outer layers) with effective convection coefficient**

In the absence of heat source in the aluminum tank and fiberglass layers, the total heat flux $\bar{Q}^{conv} = 2\pi r_1 H \bar{q}^{conv}$ at the soil surface $r = r_1$ is equal to the total heat flux $Q^{conv} = 2\pi R_o H q^{conv}$ at the outermost convection boundary $r = R_o$, where $H$ is the height of cylinder. Therefore, the heat flux density at the soil surface can be represented as:

$$\bar{q}^{conv} = \frac{R_o}{r_1} q^{conv}. \tag{32}$$

Also, in each outer layer, the temperature gradient can be written as:

$$-2\pi r H k^{Al} T_{,r} = 2\pi R_o H q^{conv} \rightarrow T_{,r} = -\frac{q^{conv}}{k^{Al}} \frac{1}{r/R_o}, \tag{33}$$
$$\text{for } r \in (r_1, r_2],$$

$$-2\pi r H k^{FG} T_{,r} = 2\pi R_o H q^{conv} \rightarrow T_{,r} = -\frac{q^{conv}}{k^{FG}} \frac{1}{r/R_o}, \tag{34}$$
$$\text{for } r \in (r_2, R_o].$$

Then, the temperature at the outermost boundary $\bar{T}$ can be written as:

$$\bar{T} = \bar{T}^s + \int_{r_1}^{r_o} T_{,r} \, dr = \bar{T}^s - q^{conv} R_o A, \tag{35}$$

where

$$A \equiv \frac{\log(r_2/r_1)}{k^{Al}} + \frac{\log(r_o/r_2)}{k^{FG}}. \tag{36}$$

The convection boundary of the full model can be reformulated as:

$$q^{conv} = c(\bar{T} - T^{env}) = c(\bar{T}^s - q^{conv} R_o A - T^{env}), \tag{37}$$

which leads to

$$q^{conv} = \frac{c}{1 + c R_o A}(\bar{T}^s - T^{env}). \tag{38}$$

Therefore, the convection boundary of the reduced model can be formulated as:

$$\bar{q}^{conv} = \frac{R_o}{r_1} q^{conv} = \frac{R_o}{r_1} \frac{c}{1 + c R_o A}(\bar{T}^s - T^{env}) \equiv \bar{c}(\bar{T}^s - T^{env}), \tag{39}$$

where the effective convection coefficient $\bar{c}$ is defined as:

$$\bar{c} = \frac{R_o}{r_1} \frac{c}{1 + c R_o A}. \tag{40}$$

### 4.3 SCNI under axisymmetry

Due to axisymmetry in the experimental tank scale test, the SCNI smoothed gradients in Eq. (27) are modified as follows to account for the cylindrical coordinates:

$$\begin{aligned}
\widetilde{\Psi}_{I,r}(\mathbf{x}_L) &= \frac{1}{r_L W_L} \int_{\Omega_L} \Psi_{I,r}(\mathbf{x}) r \, d\Omega \\
&= \frac{1}{r_L W_L} \left( \int_{\Gamma_L} \Psi_I(\mathbf{x}) n_r r \, d\Omega - \int_{\Omega_L} \Psi_I(\mathbf{x}) \, d\Omega \right),
\end{aligned} \tag{41}$$

$$\widetilde{\Psi}_{I,z}(\mathbf{x}_L) = \frac{1}{r_L W_L} \int_{\Omega_L} \Psi_{I,z}(\mathbf{x}) r \, d\Omega = \frac{1}{r_L W_L} \int_{\Gamma_L} \Psi_I(\mathbf{x}) n_z r \, d\Gamma, \tag{42}$$

where $\widetilde{\Psi}_{I,r}(\mathbf{x}_L)$ and $\widetilde{\Psi}_{I,z}(\mathbf{x}_L)$ are the smoothed gradients of $\Psi_I$ evaluated at $\mathbf{x}_L$ nodal integration point with respect to $r$ and $z$ directions, respectively.

To verify the modified SCNI for axisymmetry, an axisymmetric linear patch test is performed with a linear manufactured solution: $u_r = 0.1r$ and $u_z = 0.2z$. As shown in Figure 4-4, the modified axisymmetric SCNI can reproduce linear displacement fields and constant strain fields exactly. However, Figure 4-5 illustrates that the unmodified SCNI yields large errors, especially in strain fields.

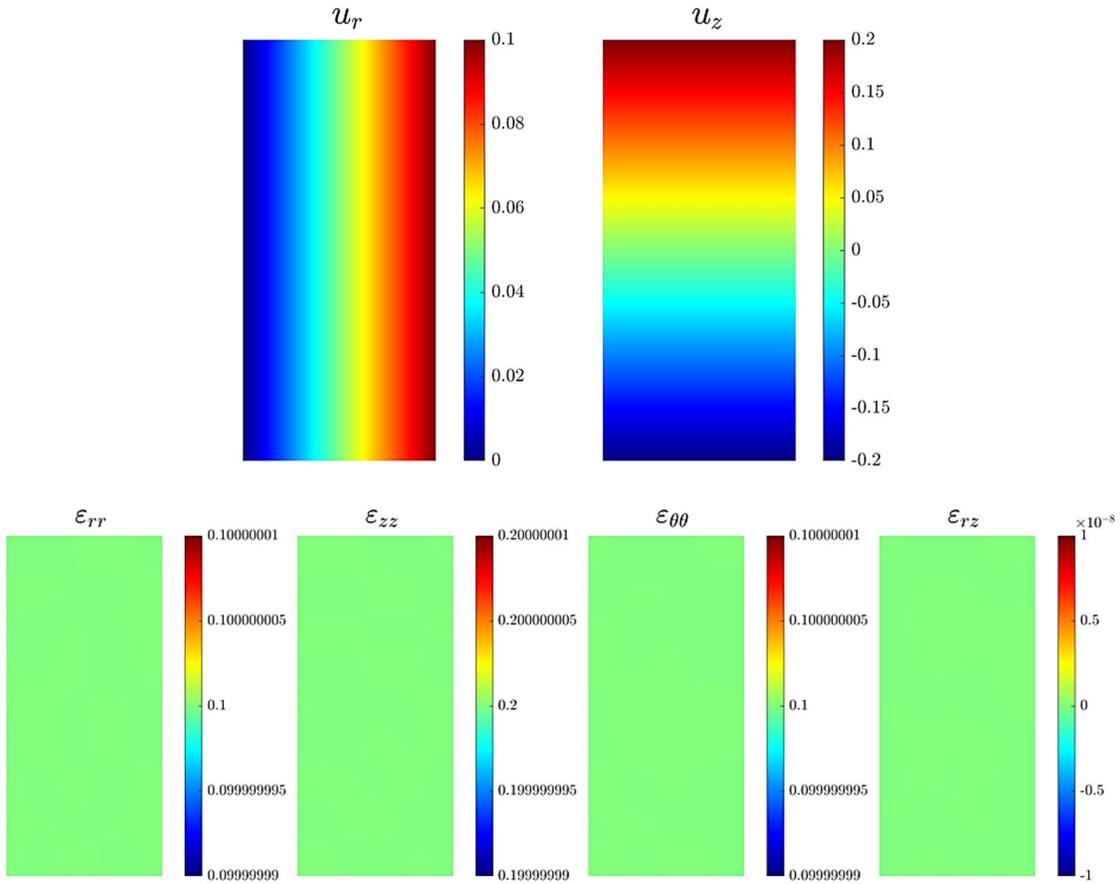

**Figure 4-4: Numerical solution obtained by the modified axisymmetric SCNI**

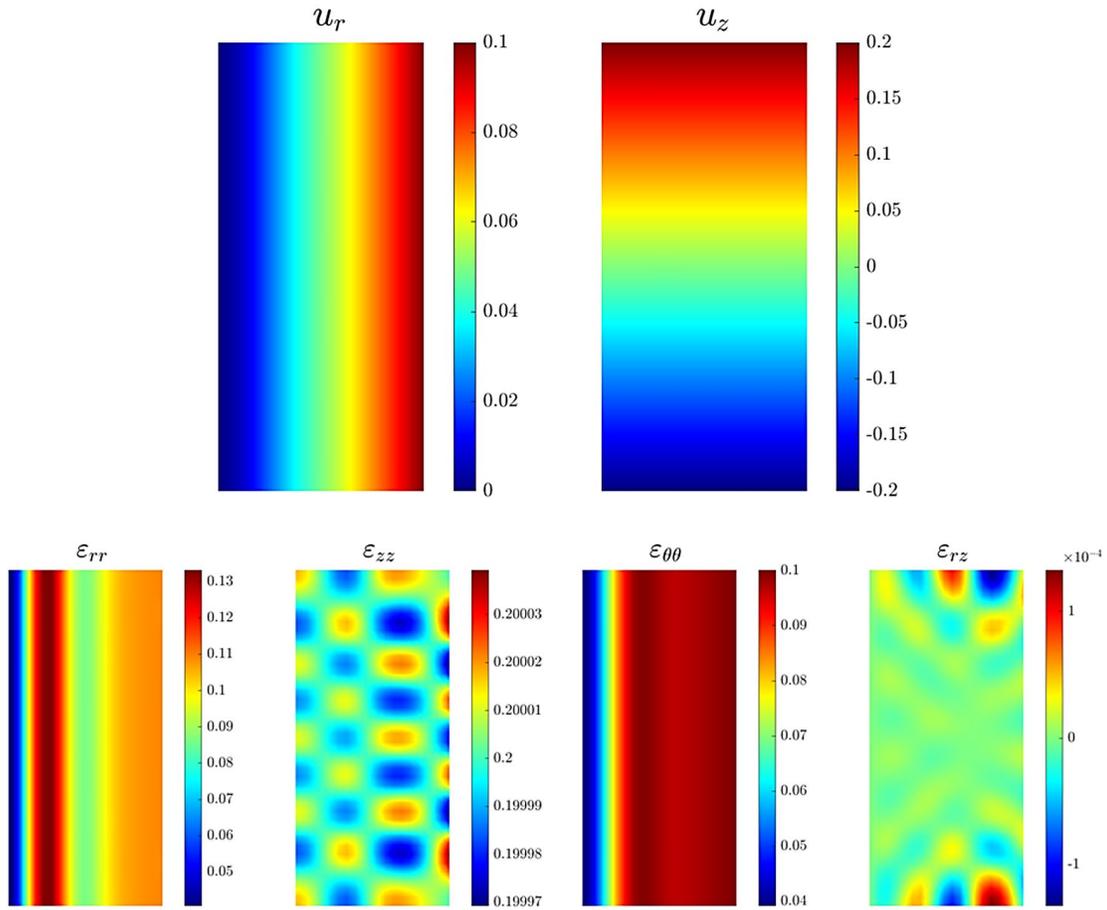

**Figure 4-5: Numerical solution obtained by unmodified SCNI**

## 5 Data-driven Modeling of Soil Water Retention Curve

A data-driven computing approach is performed to identify the THM constitutive relationships for compacted MX-80 bentonite from the experimental measurements. Figure 5-1 shows the continuum THM modeling hierarchy and the connections between the THM variables in the constitutive laws and balance equations. While some equations, such as the balance equations and Darcy's law, are physics-based, the THM material behaviors and the path-dependent effective stress constitutive laws are often partially or fully phenomenological. For example, the available constitutive models for the effects of temperature and volume change on the THM behavior of bentonite are fully phenomenological, including the SWRC, the HCF, the saturation-dependent TCF, and the VHCF. Traditionally, phenomenological functions are defined for each of these curves based on experimental observations, mechanistic hypotheses, and mathematical assumptions. However, limited data and functional form assumptions inevitably introduce errors to the model parameter calibration and model prediction. Further, pre-defined functions often lack generality and flexibility to capture a wide range of complex coupling behaviors.

In this section, a machine learning enhanced data-driven constitutive modeling approach is presented as an alternative of the conventional constitutive modeling, with a focus on the data-driven SWRC response. The deep neural networks (DNNs), as the core of deep learning, have been successfully applied in various domains due to their strong flexibility and capability in extracting complex features and patterns from data (Goodfellow et al. 2016). In this work, we explore the predictivity of DNN machine learning technique for modeling the thermo-hydraulic behavior of bentonite during central heating.

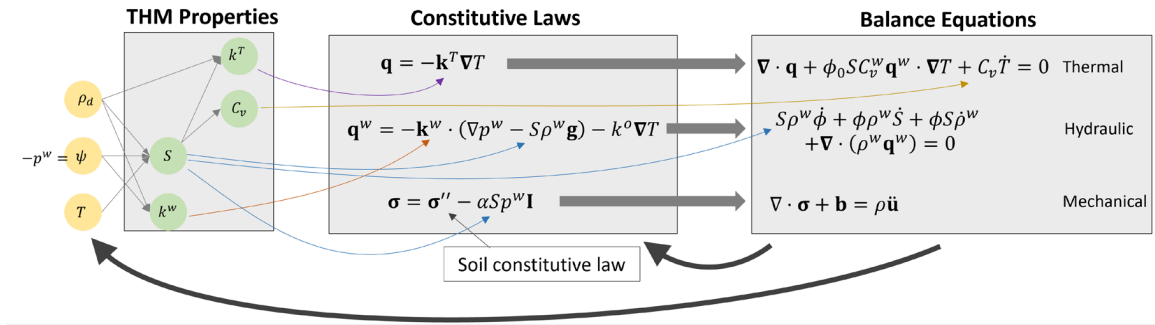

**Figure 5-1: Hierarchy and interaction of THM modeling**

## 5.1 Deep Neural Networks (DNNs)

DNNs represent complex models that relate data inputs, $\boldsymbol{x} \in \mathbb{R}^{N_{in}}$, to data outputs, $\boldsymbol{y} \in \mathbb{R}^{N_{out}}$, where $N_{in}$ and $N_{out}$ denote the number of input and output features, respectively. A typical DNN is composed of an input layer, an output layer, and a stack of fully connected hidden layers in-between, with each layer transforming the outputs of the previous layer through an affine mapping followed by a nonlinear activation function $f(\cdot)$, expressed as:

$$x^{(l)} = f(W^{(l)}x^{(l-1)} + b^{(l)}), \quad l = 1, \ldots, L, \tag{43}$$

where $L$ is the total number of hidden layers, and the superscript $(l)$ denotes the layer to which the quantities belong. For example, $x^{(l)} \in \mathbb{R}^{N_l}$ is the output of the layer $l$ with $N_l$ neurons. In addition, $W^{(l)} \in \mathbb{R}^{N_l \times N_{l-1}}$ and $b^{(l)} \in \mathbb{R}^{N_l}$ are the trainable weight matrix and the bias vector, respectively. A fully connected layer has $(N_{l-1} + 1)N_l$ trainable parameters.

Commonly used activation functions include the logistic sigmoid function, the hyperbolic tangent function, and the rectified linear unit (Xu et al. 2015). The choice of the activation of the output layer depends on the type of machine learning tasks. For regression tasks, which is the application of this work, a linear function is used in the output layer where the output from the last hidden layer, $x^{(L)}$, is mapped to the output vector $\hat{y}$, expressed as $\hat{y} = W^{(L+1)}x^{(L)} + b^{(L)}$, with $\hat{y}$ denoting the DNN approximation of the target output $y$. Figure 5-2 shows the computational graph of a fully connected feed-forward DNN with two hidden layers.

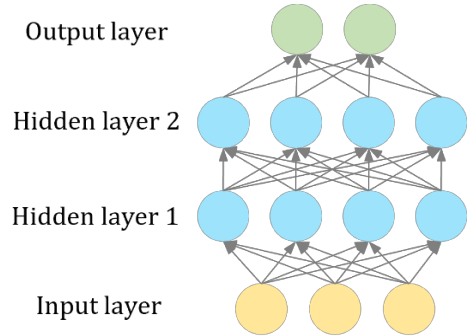

**Figure 5-2: The computational graph of a fully connected feed-forward deep neural network with two hidden layers**

## 5.2 DNN-based SWRC Constitutive Model

### 5.2.1 Training data preparation

A DNN is applied to model the non-isothermal soil water retention behaviors of MX-80 bentonite. The input variables of the DNN consist of the matric suction $\psi$ (MPa) and temperature $T$ (°C), while the output is the degree of saturation $S$ (m$^3$/m$^3$).

The training data used in this work are prepared utilizing suitable experimental data from the literature and an existing generalized isothermal SWRC phenomenological equation. Villa and Gómez-Espina (Villar and Gómez-Espina 2008; Villar and Gòmez-Espina 2007) conducted a series of experiments using the sensor/cell method for MX-80 bentonite compacted at a dry density of 1.60 g/cm³ to determine the average equilibrium values of matric suction at various temperatures. As illustrated in Figure 5-3, for each gravimetric water content level ($w$), 4 to 6 pairs of suction-temperature data points are provided from the experiments of Villa and Gómez-Espina (Villar and Gómez-Espina 2008; Villar and Gòmez-Espina 2007).

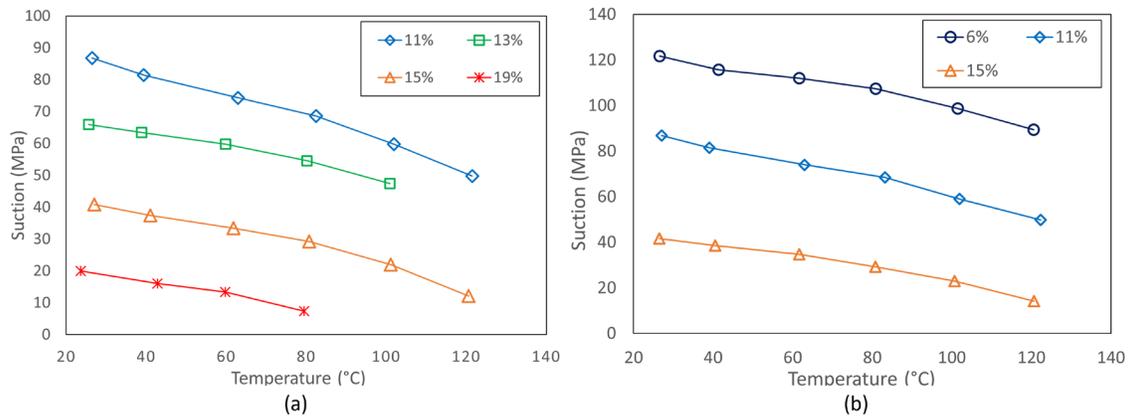

Figure 5-3: Equilibrium suction measured during heating in blocks of MX-80 bentonite compacted at $1.60\ g/cm^3$ with different gravimetric water contents: (a) recreated from Figure 3 in (Villar and Gòmez-Espina 2007) and (b) recreated from Figure 9 in (Villar and Gómez-Espina 2008)

As shown in Figure 5-3, the data points available from these experiments are very limited and sparse, consisting of information at only 5 water content levels, which correspond to 5 distinct degrees of saturation. To effectively train a generalizable DNN-SWRC constitutive model that learns complicated non-isothermal soil water retention behaviors, we enrich the training data set by a generalized isothermal soil water retention equation proposed by Lu (Lu 2016), which is shown to be a statistical improvement over other existing phenomenological SWRC models in both high and full suction ranges. The SWRC model of Lu is summarized as follows:

$$S(\psi) = \frac{\theta(\psi)}{\phi} = [\theta_a(\psi) + \theta_c(\psi)]/\phi,$$

$$\theta_a(\psi) = \theta_{a,max}\left\{1 - \left[\exp\left(\frac{\psi - \psi_{max}}{\psi}\right)\right]^m\right\}, \quad (44)$$

$$\theta_c(\psi) = \frac{1}{2}\left[1 - \text{erf}\left(\sqrt{2}\frac{\psi - \psi_c}{\psi_c}\right)\right][\theta_s - \theta_a(\psi)][1 + (\alpha\psi)^n]^{\frac{1}{n}-1},$$

where $\theta$ is the volumetric water content, $\theta_a$ and $\theta_c$ are the adsorbed water and capillary water under the prevailing suction $\psi$, respectively, and erf(·) is the Error function. Note that Lu's model involves 7 fitting parameters. For the conditions of the tank-scale test described in Section 2, the saturated volumetric water content $\theta_s$ is equal to the known soil sample porosity $\phi = 0.42$ based on a dry density of 1.60 g/cm³ and a specific gravity $G_s = 2.76$. Additionally, the adsorption strength $m$ can be commonly assumed to be related to the capillary pore-size distribution $n$ (Genuchten 1980) as $m = 1 - 1/n$, which reduces the number of unknown parameters to five. The flexibility of the SWRC equation of Lu is well demonstrated in (Lu 2016) as it can be calibrated for a large variety of soil compositions. However, Lu's SWRC equation is isothermal so that one set of fitted parameters are not generalizable to represent one type of soil at different temperatures. Therefore, individual curves at different temperatures are fitted using Lu's isothermal phenomenological SWRC equation with available experimental data points in Figure 5-3, in order to enrich the training data set for the DNN model to learn the overall soil water retention behaviors under non-isothermal conditions.

Given a temperature, $\psi - S$ data points can be sampled from Villa's and Gómez-Espina's experimental results (Villar and Gómez-Espina 2008; Villar and Gòmez-Espina 2007), as shown in Figure 5-3, which are used to fit Lu's isothermal SWRC equation for $T = 24, 30, 40, 50, 60, 70, 80, 90$ and $100\,°C$, respectively. Note that for the 19% gravimetric water content curve, the suction data at 90 and 100 °C are interpolated using a third order polynomial to provide sufficient data pairs for fitting Lu's model. At each temperature, five unknown parameters of are determined through the non-linear least-square method with a trust-region optimization algorithm (Byrd et al. 1988). The final training data set contains 2,671 training data points from both the experimental data and the fitted Lu's isothermal SWRC equations at various temperatures, as shown in Figure 5-4.

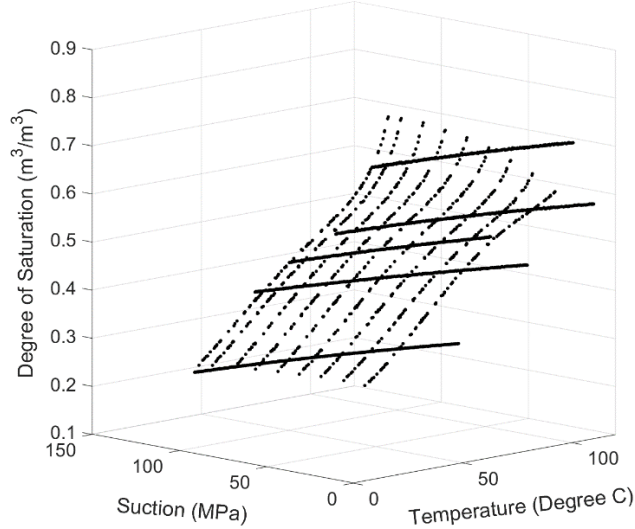

**Figure 5-4: Distribution of the training data points**

To consider the uncertainty nature of data and enhance the model performance, stochasticity is introduced to the training data points shown in Figure 5-4. Random perturbations generated from a normal (Gaussian) distribution with a zero mean and standard deviations of $rT_{max}$, $r\psi_{max}$, and $rS_{max}$ for $T$, $\psi$, and $S$ data, respectively, are added to the training data, where $T_{max}$, $\psi_{max}$, and $S_{max}$ are maximum temperature, suction, and degree of saturation in the training data, and $r$ is a user-defined parameter to control the level of randomness. For the current work, $r$ is chosen to be 0.1%. The ranges of final training inputs $T$ and $\psi$, and training output $S$ are $T \in [24, 120\ °C]$, $\psi \in [0.23, 121.34\ \text{MPa}]$, and $S \in [0.14, 0.84]$, respectively.

### 5.2.2 DNN-SWRC model training

Figure 5-5 demonstrates the computational graph of the DNN for modeling soil water retention behaviors. The proposed DNN-SWRC has two hidden layers with 100 neurons. The total 2,671 data points are partitioned into a training set and a testing set, where 80% of data are used for model training, and the rest of data are used for validation. To accelerate the training process, the training dataset is standardized to have zero mean

and unit variance. For instance, the input temperature data are standardized by the mean $\mu_T$ and standard deviation $std_T$:

$$\bar{T} = S_T(T) = \frac{T - \mu_T}{std_T}. \tag{45}$$

The same standardization operation is applied to the input $\psi$ and ground truth output $S$, and hereafter, an overbar symbol denotes standardized quantities.

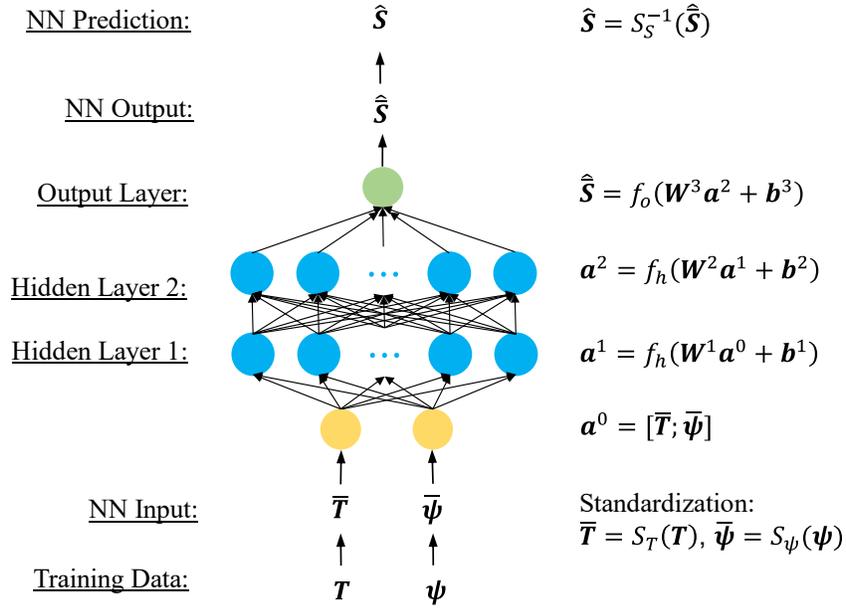

Figure 5-5: The computational graph of the DNN-SWRC model

The training consists of forward and backward propagations. During the forward propagation, a series of operations defined in Eq. (43) are carried out through layers, which are illustrated in Figure 5-5, where $f_h$ and $f_o$ denote the activation functions used for the hidden layer and the output layer, respectively, and $\boldsymbol{W}^l$ and $\boldsymbol{b}^l, l = 1,2,3$, are model's trainable parameters. The hyperbolic tangent function is utilized as the hidden layer activation function, and linear activation is adopted for the output layer. Note that the target output data is standardized and therefore the DNN output is considered as standardized. The DNN predicted degrees of saturation $\widehat{S}$ can be obtained as follows:

$$\widehat{\boldsymbol{S}} = \widehat{\bar{\boldsymbol{S}}} \cdot std_S + \mu_s \equiv S_S^{-1}\left(\widehat{\bar{\boldsymbol{S}}}\right). \tag{46}$$

For simplicity, $\boldsymbol{\Theta} = \{\boldsymbol{W}, \boldsymbol{b}\}$ is used to denote all trainable parameters of the DNN, where $\boldsymbol{W} = \{\boldsymbol{W}^{(l)}\}_{l=1}^{3}$ and $\boldsymbol{b} = \{\boldsymbol{b}^{(l)}\}_{l=1}^{3}$, and let $\boldsymbol{x} = \{\boldsymbol{T}, \boldsymbol{\psi}\}$ represents the inputs of DNN. The backward propagation updates the trainable parameters by the gradient of the loss function. The optimal parameters of the DNN are obtained by:

$$\boldsymbol{\Theta}^* = \arg\min_{\boldsymbol{\Theta}} \left\{ \sum_{i=1}^{N_s} \left[ \left\| \bar{S}_i - \hat{\bar{S}}_i(\boldsymbol{x}_i; \boldsymbol{\Theta}) \right\|_{L_2}^2 + \beta_1 \text{ReLU}(\hat{S}_i(\boldsymbol{x}_i; \boldsymbol{\Theta}) - 1) \right] + \beta_2 \sum_{l=1}^{3} \left\| \boldsymbol{W}^{(l)} \right\|_F^2 \right\}. \tag{47}$$

Where $N_s$ denotes the number of training samples. The first term on the right-hand side of Eq. (47) calculates the sum of squared Euclidean distances between the standardized ground truth data and DNN outputs. The second term is to constrain the DNN predicted degrees of saturation such that $\hat{S}_i \leq 1$, which is realized through the rectified linear unit function (ReLU), and $\beta_1$ is a penalty parameter. The third term is a regularization term used to prevent over-fitting issues, where $\|\cdot\|_F$ denotes the Frobenius norm, and $\beta_2$ is the regularization parameter for the training weights. For the current model, $\beta_1$ is selected to be 1 and $\beta_2 = 10^{-5}$. The optimization is performed using the MATLAB Deep learning toolbox (MathWorks 2022), and the Adam optimizer (Kingma and Ba 2015) is adopted for the back-propagation training with an initial learning rate of $10^{-3}$.

The prediction accuracy of the trained DNN-SWRC is measured using a relative error defined as follows:

$$e = \frac{\|\boldsymbol{S} - \hat{\boldsymbol{S}}\|_{L_2}}{\|\boldsymbol{S}\|_{L_2}}. \tag{48}$$

As shown in Figure 5-6, the proposed DNN-SWRC achieves relative errors of 1.167% and 1.310% for the training and testing datasets, respectively, demonstrating effective training and accurate generalization of the DNN-SWRC model.

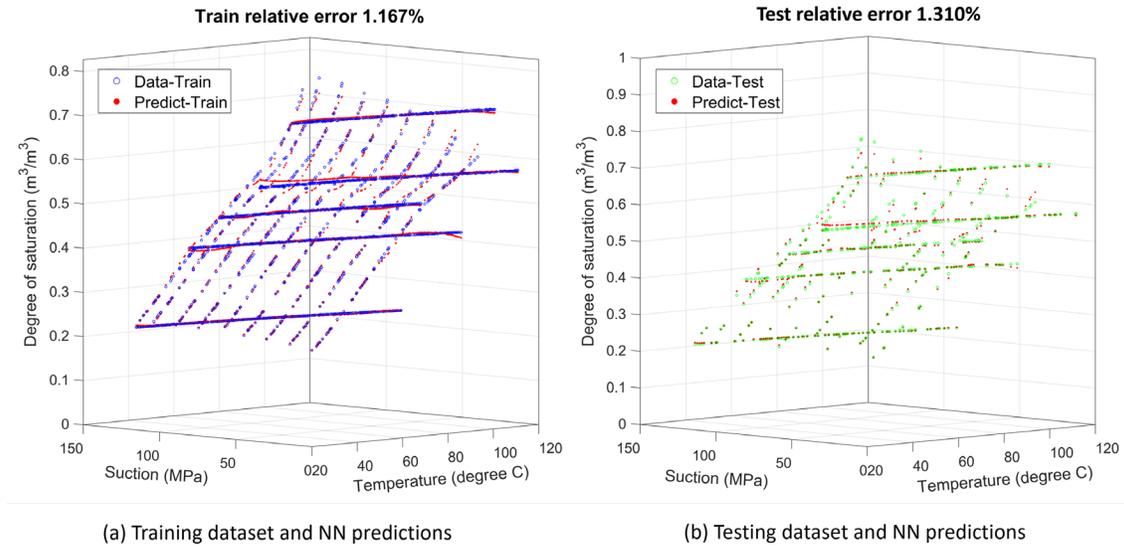

(a) Training dataset and NN predictions  (b) Testing dataset and NN predictions

**Figure 5-6: Comparison between the data and the predictions of the trained DNN-SWRC model: (a) with respect to the training dataset and (b) with respect to the testing dataset**

### 5.2.3 Comparison of DNN-SWRC and Lu's isothermal SWRC constitutive relationship

The accuracy of the proposed DNN-SWRC model is also analyzed using two sets of experiment data provided in (Villar and Gómez-Espina 2008) and (Villar and Gòmez-Espina 2007), which is compared to fitted generalized SWRC model of Lu at various temperatures. A root-mean square error (RMSE) between the predicted and measured degrees of saturation is utilized as an accuracy measurement:

$$\text{RMSE} = \sqrt{\frac{1}{N_d} \sum_{i=1}^{N_d} (S_i - \hat{S}_i)^2}, \tag{49}$$

where $S_i$ and $\hat{S}_i$ denote the experimentally measured and model predicted degrees of saturation, respectively, and $N_d$ is the total number of experiment data points at each temperature used to assess the accuracy.

The first set of data used for model accuracy assessment are those shown in Figure 5-3, which are experimental data used to fit Lu's SWRC equations for $T =$ 30, 40, 50, 60, 70, 80, 90 and 100 ℃, respectively, and are included in the training of the DNN-SWRC model. At each temperature, the trained DNN-SWRC and fitted Lu's equations are used to predict the degrees of saturation for $\psi \in [1, 120 \text{ MPa}]$ with an increment of 1 MPa. Figure 5-7 plots the DNN predicted SWRC and Lu's SWRC equations at 8 different temperatures between 30 to 100 ℃, and for each temperature, five experiment data points are also shown in Figure 5-7. Table 5-1 summarizes the RMSE values for the predictions made by the DNN-SWRC and fitted Lu's SWRC equations at different temperatures compared to Villa's and Gómez-Espina's experimental data. It shows that the trained non-isothermal DNN-SWRC model achieves higher accuracy than the fitted Lu's SWRC models at all temperatures.

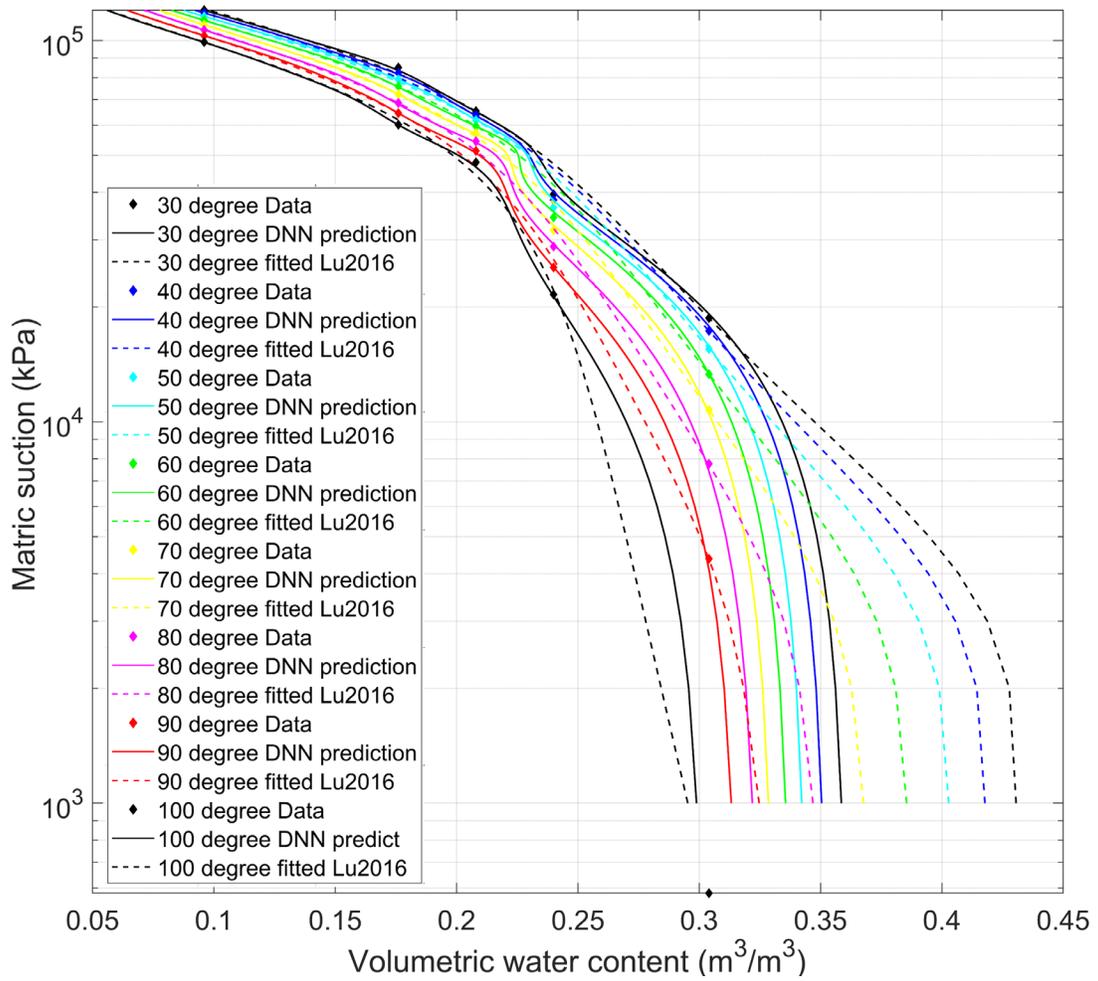

**Figure 5-7: DNN-SWRC and Lu's SWRC models at various temperatures** (compared with experiment data obtained from Figure. 3 and Figure. 9 in (Villar and Gómez-Espina 2008) and (Villar and Gòmez-Espina 2007))

| Temperature (°C) | Root-Mean Square Error (RMSE) | |
|---|---|---|
| | Fitted Lu's SWRC of Lu | DNN-SWRC |
| 100 | 3.20E-03 | 2.00E-03 |
| 90 | 3.70E-03 | 8.53E-04 |
| 80 | 3.50E-03 | 1.10E-03 |
| 70 | 4.50E-03 | 1.10E-03 |
| 60 | 5.80E-03 | 1.40E-03 |
| 50 | 6.20E-03 | 1.80E-03 |
| 40 | 6.70E-03 | 2.20E-03 |
| 30 | 7.10E-03 | 2.80E-03 |

**Table 5-1: RMSE of predictions made by DNN-SWRC and fitted Lu's SWRC models at various temperatures** (compared with experiment data obtained from Figure. 3 and Figure. 9 in (Villar and Gómez-Espina 2008) and (Villar and Gòmez-Espina 2007))

In addition to the above comparison where the experimental data are used for training the DNN-SWRC model, we further compare the trained DNN-SWRC with the Lu's SWRC models against additional unseen experimental $\psi - S$ data points during the training of the DNN-SWRC model at $T = 26, 41, 81$ and $101$ °C provided in (Villar and Gòmez-Espina 2007), as shown in Figure 5-8. The unknown parameters in Lu's isothermal SWRC models are first estimated at those four temperatures through fitting the experiment data.

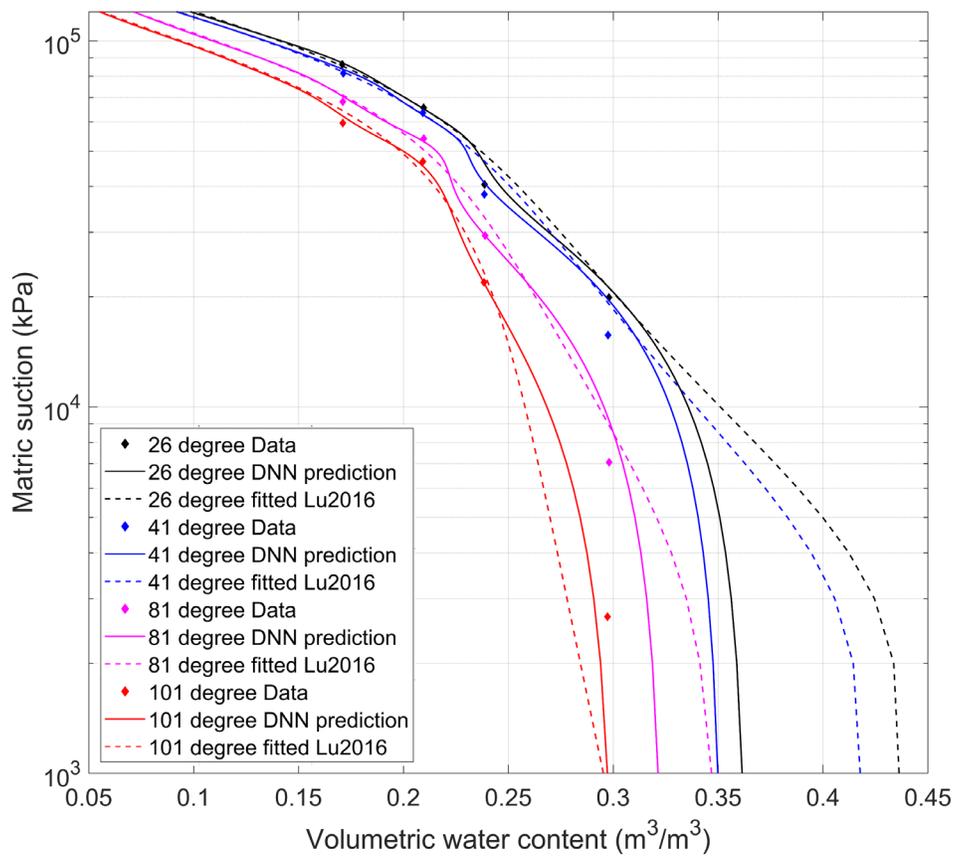

**Figure 5-8: DNN-SWRC and Lu's SWRC models at various temperatures** (compared with experiment data obtained from Figure. 5 in (Villar and Gòmez-Espina 2007))

| Temperature (°C) | Root-Mean Square Error (RMSE) | |
| --- | --- | --- |
| | Fitted SWRC of Lu | DNN-SWRC |
| 26 | 7.90E-03 | 4.30E-03 |
| 41 | 1.01E-02 | 6.90E-03 |
| 81 | 7.10E-03 | 4.30E-03 |
| 101 | 1.04E-02 | 4.40E-03 |

**Table 5-2: RMSE of predictions made by DNN-SWRC and fitted Lu's SWRC models at various temperatures** (compared with experiment data obtained from Figure. 5 in (Villar and Gòmez-Espina 2007))

Since the new experiment data points are within the training range of the DNN-SWRC model, no additional training is needed. Then, for $\psi \in [1, 120 \text{ MPa}]$, the degrees of saturations are predicted by both the trained DNN-SWRC and fitted Lu's SWRC models at $T = 26, 41, 81$ and $101 \,°\text{C}$, and the results are shown in Figure 5-8 along with the experiment data points. Table 5-2 summarizes the RMSE of the predictions from the DNN-SWRC and fitted Lu's SWRC equations at different temperatures with respect to the experimental data. It shows that the trained DNN-SWRC model exhibits higher prediction accuracy than Lu's SWRC models, even with unseen data.

# 6 Tank-scale Simulation with DNN-SWRC model

The trained DNN-SWRC model proposed in Section 5.2 is utilized in place of phenomenological equations to evaluate the degree of saturation for a given water pressure and temperature. The isotropic, saturation-dependent hydraulic conductivity $k^w$ is obtained from the van Genuchten-Mualem model (Genuchten 1980). Lu and McCartney (Lu and Mccartney 2023a) found that the shape of the HCF is primarily dependent on the shape of the SWRC in the capillary regime, which is similar to the van Genuchten SWRC in Lu's SWRC model. The saturation-dependent thermal conductivity $k^T$ and the volumetric heat capacity $C_v$ are evaluated using a phenomenological thermal conductivity function proposed by Lu and Dong (Lu and Dong 2015). Then, the proposed enhanced RK numerical model discussed in Section 3, incorporated with the evaluated thermo-hydraulic properties of bentonite, is applied to the heating stage of an axisymmetric tank scale simulation of MX80 bentonite (Lu and McCartney 2022).

## 6.1 Numerical model setups

As demonstrated in the experimental set up of the tank-scale test in Section 2, both the geometry and heating condition of the tank scale experiment are axisymmetric. Therefore, a 2D half-section of the tank is simulated, as shown in Figure 6-1. The left side of the numerical model has a prescribed temperature of $200 \,°\text{C}$ across the duration of the simulation to mimic the behavior of the center heating element. The measurements of the temperature evolutions of the top-center and side of the soil layer (Figure 6 of (Lu and McCartney 2022)) indicate slight temperature increment at the outer boundaries of the container. Therefore, convective heat transfer boundaries are applied to the top, bottom, and right sides of the numerical model with an effective convective coefficient $\bar{c} = 2 \text{ Wm}^{-2}\text{K}^{-1}$. Since there is no water inflow during the heating stage of the experiment, zero water flux is enforced at all four boundaries.

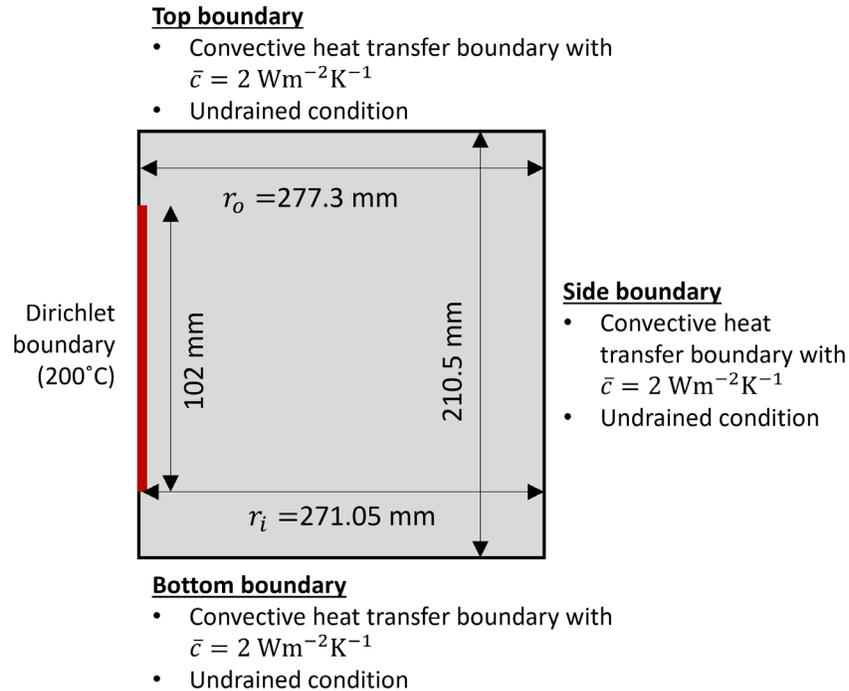

Figure 6-1: A schematic of numerical model setup

## 6.2 Numerical and experimental results

It is worth mentioning that the experimental data from (Villar and Gómez-Espina 2008; Villar and Gòmez-Espina 2007) involved in training the DNN-SWRC are obtained under a testing environment different from that of the tank scale experiment. In Villa's and Gómez-Espina's experiments (Villar and Gómez-Espina 2008; Villar and Gòmez-Espina 2007), the constant-volumed bentonite blocks with fixed gravimetric water contents were kept inside a stainless-steel hermetic cell during the experiments, and isotropic heating was applied to the cell. Under such conditions, constant water contents, which correspond to constant degrees of saturation, can be kept within the soil specimens at temperatures exceeding the water boiling points at one atmospheric pressure. However, in the tank scale experiment, no hermetic seal is involved and the water in the proximity of the center heater is expected to completely evaporate from the soil and diffuse outward, causing a temporary wetting front that passes by the sensors (Lu and McCartney 2022). As such, a hyperbolic tangent correction function is proposed in Eq. (50) to scale down the DNN-SWRC predicted degrees of saturations for the high-temperature range to account for the difference in testing condition between the training data and the actual tank scale experiment.

$$H(T) = 0.5\left(1 - \tanh\left(\frac{T-T_{cr}}{c_s}\right)\right), \tag{50}$$

where $T_{cr}$ and $c_s$ denote the center and smoothing width of the correction function. Figure 6-2 illustrates the effects of $T_{cr}$ and $c_s$ on the correction function. A scaling center of 80 °C and a smoothing width of 20 °C are selected in the current work. Accordingly, the degree of saturation used for the numerical simulation is obtained by:

$$\tilde{S} = H(T)S^{DNN-SWRC}(\psi, T). \tag{51}$$

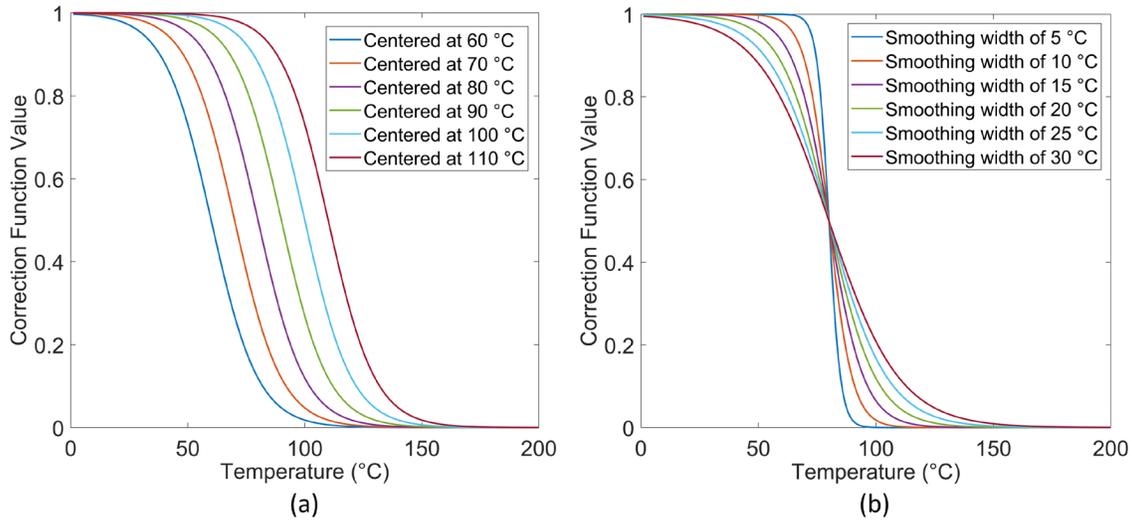

**Figure 6-2: Effects of parameters of the correction function on the function value: (a) varying scaling centers with a fixed smoothing width of 20 °C and (b) varying scaling widths with a fixed scaling center of 80 °C**

Figure 6-3 compares the numerical results of the tank scale simulation with the experimentally data measured at five sensor locations. The numerically predicted temperature history at all five sensor locations agrees well with the experimental data,

showing significant temperature increases at all sensor locations within the first 100 hours of heating and then gradually reaching equilibrium, as illustrated in Figure 6-3 (a). Figure 6-3 (b) compares the time histories of degrees of saturation obtained from the numerical simulations with the experimental data. The predicted degrees of saturation exhibit an initial increase of the water content in the bentonite, followed by a gradual drying corresponding to a decrease of soil's degree of saturation, which shows a good agreement with the experimental observations.

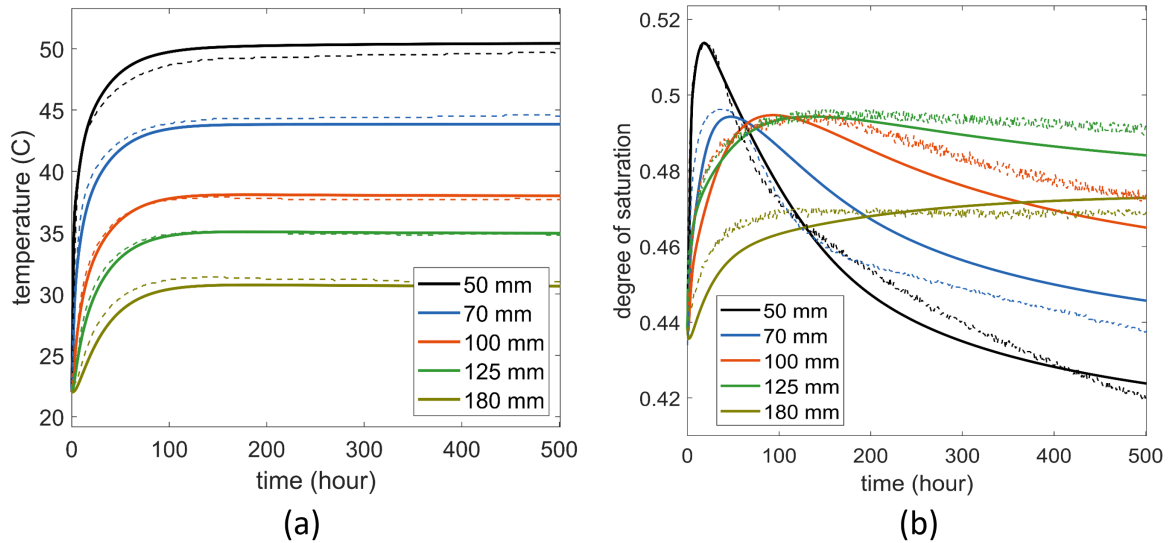

**Figure 6-3: Comparison between numerical simulation results of the tank scale experiment with the experimental data: (a) temperature results (b) degree of saturation results** (solid lines: numerical, dashed lines: experimental)

# 7   Conclusion

A meshfree numerical framework based on the reproducing kernel particle method is proposed in this work for modeling the coupled heat transfer and hydrations of MX-80 bentonite buffers. An enrichment basis based on the derived analytical solution of a radial plane-strain heat conduction boundary value problem is introduced in the approximation of the temperature field to effectively capture high temperature gradients near the heat source. No additional degree of freedom is required for the proposed enrichment, and its effectiveness is demonstrated through a radial-plane strain heat transfer problem. For improved numerical efficiency and stability, the SCNI is reformulated for axisymmetric geometries, ensuring exact integration of linear axisymmetric solution fields. A semi-analytical treatment is employed for convection boundaries involving thin composite layers, which introduces an effective heat transfer coefficient and eliminates the need for modeling outer thin composite layers, thereby easing constraints on nodal spacing.

Additionally, this work proposes a non-isothermal, data-driven, DNN-based SWRC constitutive model that takes temperature and matric suction as inputs and predicts the degree of saturation. The training of the DNN-SWRC utilizes experimental data from literature and fitted Lu's SWRC equation at different temperatures. Compared to the conventional phenomenological SWRC models, the trained DNN-SWRC model demonstrates higher accuracy and applicability in representing the water retention behaviors of MX-80 bentonite for wide ranges of temperatures and hydrations. Through validation against experimental measurements and comparison with the fitted Lu's SWRC model, the DNN-SWRC model exhibits superior predictive accuracy.

The trained DNN-SWRC model is incorporated into the proposed meshfree numerical scheme as the constitutive relation for water retention, which is employed to model a tank-scale heating experiment. To accommodate the discrepancies between the experimental conditions of the training dataset for DNN-SWRC and those of the tank-scale experiments, correction based on a hyperbolic tangent function is performed to scale down the DNN-SWRC predicted degrees of saturation in the higher temperature range. A good agreement between the numerical results and experimental measurements is achieved, in terms of temporal evolutions of the temperature, suction, and the degree of saturation at five sensor locations, underlining the efficacy of the proposed data-driven numerical framework. Note that current work focuses on numerical simulations of the heating phase of tank-scale experiments. Future work will involve developing a fully coupled thermo-hydro-mechanical meshfree framework for modeling the hydration phase. Additionally, data-driven machine learning models will be extended to consider other important thermohydraulic properties of bentonite.

## Acknowledgements

The authors appreciate support from US Department of Energy Nuclear Energy University Program award DE-NE008951. The views in this paper are those of the authors alone.

Characteristic Curve Using Genetic Programming." *Journal of Geotechnical and Geoenvironmental Engineering* 132 (5): 661–65. https://doi.org/10.1061/(ASCE)1090-0241(2006)132:5(661)/ASSET/0EB1E28E-C3C4-4C1D-BCE2-59A67065949F/ASSETS/IMAGES/6.JPG.

Kadeethum, T., O'Malley, D., Fuhg, J. N., Choi, Y., Lee, J., Viswanathan, H. S., and Bouklas, N.. 2021. "A Framework for Data-Driven Solution and Parameter Estimation of PDEs Using Conditional Generative Adversarial Networks." *Nature Computational Science 2021 1:12* 1 (12): 819–29. https://doi.org/10.1038/s43588-021-00171-3.

Kaneko, S., Wei, H., He, Q., Chen, J. S., and Yoshimura, S.. 2021. "A Hyper-Reduction Computational Method for Accelerated Modeling of Thermal Cycling-Induced Plastic Deformations." *Journal of the Mechanics and Physics of Solids* 151 (June): 104385. https://doi.org/10.1016/J.JMPS.2021.104385.

Kim, Y., Choi, Y., Widemann, D., and Zohdi, T.. 2022. "A Fast and Accurate Physics-Informed Neural Network Reduced Order Model with Shallow Masked Autoencoder." *Journal of Computational Physics* 451 (February): 110841. https://doi.org/10.1016/J.JCP.2021.110841.

Kingma, D. P., and Ba, J.. 2015. "Adam: A Method for Stochastic Optimization." In *3rd International Conference on Learning Representations, ICLR 2015*. https://arxiv.org/abs/1412.6980v9.

Kirchdoerfer, T., and Ortiz, M.. 2016. "Data-Driven Computational Mechanics." *Computer Methods in Applied Mechanics and Engineering* 304 (June): 81–101. https://doi.org/10.1016/J.CMA.2016.02.001.

Koekkoek, E. J. W., and Booltink, H.. 1999. "Neural Network Models to Predict Soil Water Retention." *European Journal of Soil Science* 50 (3): 489–95. https://doi.org/10.1046/J.1365-2389.1999.00247.X.

Liu, W. K., Jun, S., and Zhang, Y. F.. 1995. "Reproducing Kernel Particle Methods." *International Journal for Numerical Methods in Fluids* 20 (8–9): 1081–1106.

https://doi.org/10.1002/FLD.1650200824.

Liu, Z., Bessa, M. A., and Liu, W. K.. 2016. "Self-Consistent Clustering Analysis: An Efficient Multi-Scale Scheme for Inelastic Heterogeneous Materials." *Computer Methods in Applied Mechanics and Engineering* 306 (July): 319–41. https://doi.org/10.1016/J.CMA.2016.04.004.

Lloret, A., Villar, M. V., Sànchez, M., Gens, A., Pintado, X., and Alonso, E. E.. 2003. "Mechanical Behaviour of Heavily Compacted Bentonite under High Suction Changes." *Geotechnique* 53 (1): 27–40. https://doi.org/10.1680/GEOT.2003.53.1.27/ASSET/IMAGES/SMALL/GEOT53-027-F27.GIF.

Lu, N.. 2016. "Generalized Soil Water Retention Equation for Adsorption and Capillarity." *Journal of Geotechnical and Geoenvironmental Engineering* 142 (10): 04016051.

Lu, N., and Dong Y.. 2015. "Closed-Form Equation for Thermal Conductivity of Unsaturated Soils at Room Temperature." *Journal of Geotechnical and Geoenvironmental Engineering* 141 (6). https://doi.org/10.1061/(ASCE)GT.1943-5606.0001295.

Lu, Y., and Mccartney, J. S.. 2023a. "Insights into the Thermo-Hydraulic Properties of Compacted MX80 Bentonite during Hydration under Elevated Temperature." *Canadian Geotechnical Journal*, June. https://doi.org/10.1139/CGJ-2022-0537.

Lu, Y., and McCartney, J. S.. 2022. "Physical Modeling of Coupled Thermohydraulic Behavior of Compacted MX80 Bentonite during Heating." *Geotechnical Testing Journal* 45 (6): 1108–26. https://doi.org/10.1520/GTJ20220054.

Lu, Y., and McCartney, J. S.. 2023b. "Thermal Conductivity Function for Unsaturated Soils Linked with Water Retention by Capillarity and Adsorption." *Journal of Geotechnical and Geoenvironmental Engineering*.

MathWorks, Inc. 2022. "Deep Learning Toolbox: User's Guide (R2022b)."

McCartney, J. S., and Baser, T.. 2017. "Role of Coupled Processes in Thermal Energy Storage in the Vadose Zone." *2nd Symposium on Coupled Phenomena in*
46

Formulation for Fully Coupled Hydro-Mechanical Analysis of Fluid-Saturated Porous Media." *Computers & Fluids* 141 (December): 105–15. https://doi.org/10.1016/J.COMPFLUID.2015.11.002.

Wei, H., Wu, C. T., Hu, W., Su, T. H., Oura, H., Nishi, M., Naito, T., Chung, S., and Shen, L.. 2023. "LS-DYNA Machine Learning–Based Multiscale Method for Nonlinear Modeling of Short Fiber–Reinforced Composites." *Journal of Engineering Mechanics* 149 (3): 04023003. https://doi.org/10.1061/JENMDT.EMENG-6945/ASSET/A8C4CF4D-3014-45EE-BCAD-6A9B0FE0B0EA/ASSETS/IMAGES/LARGE/FIGURE18.JPG.

Xie, Y., and Wang, G.. 2014. "A Stabilized Iterative Scheme for Coupled Hydro-Mechanical Systems Using Reproducing Kernel Particle Method." *International Journal for Numerical Methods in Engineering* 99 (11): 819–43. https://doi.org/10.1002/NME.4704.

Xiong, Z., Xiao, M., Vlassis, N., and Sun, W. C.. 2023. "A Neural Kernel Method for Capturing Multiscale High-Dimensional Micromorphic Plasticity of Materials with Internal Structures." *Computer Methods in Applied Mechanics and Engineering* 416 (November): 116317. https://doi.org/10.1016/J.CMA.2023.116317.

Xu, B., Wang, N., Chen, T., and Li, M.. 2015. "Empirical Evaluation of Rectified Activations in Convolutional Network," *arxiv preprint* (May). https://arxiv.org/abs/1505.00853v2.

Xu, K., Huang, D. Z., and Darve, E.. 2021. "Learning Constitutive Relations Using Symmetric Positive Definite Neural Networks." *Journal of Computational Physics* 428 (March): 110072. https://doi.org/10.1016/J.JCP.2020.110072.

Zhang, J., Yang, S., Zhang, L. L., and Zhou, M. L.. 2022. "Bayesian Estimation of Soil-Water Characteristic Curves." *Canadian Geotechnical Journal* 59 (4): 569–82. https://doi.org/10.1139/CGJ-2021-0070/ASSET/IMAGES/CGJ-2021-0070TABA1.GIF.

Zheng, L., Rutqvist, J., Birkholzer, J. T., and Liu, H. H.. 2015. "On the Impact of